\documentclass[ASNA,twocolumn]{USG}
\usepackage{NJDnatbib,lettersp}
\articletype{RESEARCH ARTICLE}
\journal{}
\historydates{}
\volume{}
\copyyear{}
\startpage{1}
\articledoi{}

\makeatletter
\renewcommand{\@@address}[2][]{%
  \g@addto@macro\jmkaddress{\hsize=\textwidth
    \def\baselinestretch{1}\stepcounter{affn}%
    \xdef\@currentlabel{\theaffn}\jmkLabel{#1}%
    \addressfont\ifx#1\empty\else\@hangfrom{\textsuperscript{\smash{\theaffn}}}\fi
    #2\hphantom{1}%
    \ifnum\value{affn}=3\else\hskip-3pt\authorsep\fi
  }%
}

\def\oddfoot@titlepage@info{%
  \vbox{\hsize\textwidth
    \rule{\textwidth}{.5pt}\par\vskip3pt
    \noindent{\fontsize{7}{8}\selectfont\itshape\@journal}%
    \hfill{\pagenumfont\thepage\ of \@LastPg}\par
  }%
}
\def\evenfoot@titlepage@info{\oddfoot@titlepage@info}
\def\evenhead@headings@info{%
  \head@headings@def{\evenpage@headings@def\hfill{\pagenumfont{\itshape\@journal}}}%
}
\def\head@headings@def#1{%
  \vbox{\hsize\textwidth
    \vskip3.5mm
    \vrule width \textwidth height .5pt depth 0pt
    \vskip0pt
    \hbox to \textwidth{#1\dummyalphabet}%
  }%
}
\def\oddhead@titlepage@info{}
\def\evenhead@titlepage@info{}
\makeatother

\usepackage{placeins}
\hypersetup{hidelinks}
\RemoveFromHook{shipout/after}[USG]
\makeatletter
\newenvironment{IFEplacedfigure}{\par\addvspace{6pt}\noindent\begin{minipage}{\textwidth}\def\@captype{figure}\centering}{\end{minipage}\par\addvspace{6pt}}
\newenvironment{IFEplacedtable}{\par\addvspace{6pt}\noindent\begin{minipage}{\textwidth}\def\@captype{table}\centering\small}{\end{minipage}\par\addvspace{6pt}}
\newenvironment{IFEcolumnfigure}{\par\addvspace{5pt}\noindent\begin{minipage}{\linewidth}\def\@captype{figure}\centering}{\end{minipage}\par\addvspace{5pt}}
\makeatother
\raggedcolumns
\begin{document}
\onecolumn
\title{Enabling Mie Resonances for Optomagnetism in Epitaxially Grown Garnet Nanodisks via Photonic Crystals}
\author[1,2]{P. V. Zorina}
\author[1,3]{D. O. Ignatyeva}
\author[1]{D. M. Krichevsky}
\author[1,3]{V. I. Belotelov}
\authormark{Zorina et al.}
\titlemark{}
\address[1]{Russian Quantum Center, Moscow, 121205 Russia}
\address[2]{Moscow Institute of Physics and Technology, Dolgoprudny, 141700 Russia}
\address[3]{Faculty of Physics, M. V. Lomonosov Moscow State University, Moscow, Russia}
\abstract[ABSTRACT]{
An all-garnet photonic-crystal substrate compatible with epitaxial growth on gadolinium gallium garnet is proposed for controlling the optomagnetic response of a single bismuth iron garnet nanodisk. The photonic environment modifies the spectral positions and relative strengths of the electric- and magnetic-dipole contributions and redistributes the optical field inside the disk. Compared with a reference nanodisk on a homogeneous substrate evaluated near the broad maximum of its summed multipolar response, the peak axial optical contribution to the inverse Faraday effect in the examined cross-sections is enhanced by factors of approximately $7.3$ and $6.3$ in the magnetic- and electric-dipole-dominated regimes, respectively. The enhancement reaches approximately $7.9$ at the maximum of the summed multipolar scattering contributions. These results demonstrate spectral control of the magnitude and spatial distribution of the local optomagnetic response through the photonic environment, without changing the nanodisk geometry.
}
\maketitle
\begin{multicols}{2}
\section{Introduction}

Optomagnonic systems rely on the interaction between optical fields and collective spin excitations. They provide a promising route toward optical control and readout of magnons~\cite{wang2022recent}. The efficiency of this interaction depends strongly on the quality factor of the optical mode and on the concentration of electromagnetic energy within the magnetic material~\cite{zhang2016optomagnonic,zhu2022inverse}. Therefore, increasing the electromagnetic energy stored inside the magnetic volume is a key design requirement for optomagnonic devices. Achieving this enhancement without introducing additional optical or magnetic loss channels is essential for the development of compact resonant structures for magneto-optical and optomagnetic applications~\cite{maccaferri2020nanoscale,voronov2020magneto,rameshti2022cavity}.

Magnetic garnets are particularly suitable for optomagnonic cavities because they combine low magnetic damping, low optical losses, and pronounced magneto-optical activity. Yttrium iron garnet is commonly regarded as a benchmark material for magnonic and optomagnonic studies, while different substitutions provide an additional degree of freedom for tailoring both the optical and magnetic properties of iron garnets. In particular, increasing the Bi content enhances the magneto-optical response in the visible spectral range~\cite{jesenska2016optical}, whereas Ce-substituted YIG provides a large magneto-optical figure of merit in the near-infrared range and has been employed in integrated nonreciprocal photonic devices~\cite{sun2015single}. Importantly, such substitutions can also be used to tune the lattice parameter of the magnetic garnet and thereby achieve epitaxial matching to different garnet substrates. For example, Bi-substituted YIG can be lattice-matched to diamagnetic YSGG while retaining favorable magneto-optical and magnetic-resonance properties~\cite{kupchinskaya2024magneto,legrand2026lattice}. More generally, epitaxial growth on lattice-matched garnet substrates provides a route to high-quality, low-damping YIG and substituted-YIG films~\cite{flajvsman2026magnetic,dubs2017sub}. From the optical point of view, different garnet substrates may have very similar refractive indices and therefore provide comparably modest refractive-index contrast with magnetic iron garnets, as demonstrated, for example, for GGG and GSGG~\cite{hoefer1988thermo}. Their magnetic properties, however, can differ substantially: Gd-containing substrates can contribute additional magnetic losses, whereas the use of Gd-free diamagnetic garnets or intermediate diamagnetic garnet layers can suppress substrate-induced damping while preserving epitaxial compatibility~\cite{kuzmichev2026micrometer}. At the same time, the resonator geometry provides an additional means of strengthening light--magnon interaction by improving optical-field confinement and reducing the optical and magnon mode volumes~\cite{demirchyan2026magnon}.

Resonant enhancement of magneto-optical effects has been demonstrated for dielectric nanoantennas supporting Mie resonances~\cite{barsukova2017magneto}. Resonant magneto-optical response has also been predicted for subwavelength dielectric nanowire gratings~\cite{marinchio2014magneto} and demonstrated in all-dielectric resonant gratings~\cite{gamet2017enhancement}. All-dielectric magneto-optical metasurfaces have been used to enhance Faraday rotation in the vicinity of electric- and magnetic-dipole Mie resonances and to provide additional control of the magneto-optical response~\cite{christofi2018giant}. Related Mie-resonant metasurfaces have also been shown to enhance magnetization-induced transmission modulation and to exploit anapole-related spectral features while maintaining high transmission~\cite{zorina2026enhanced}. Further enhancement has been predicted in structures exploiting high-Q and bound-state-in-the-continuum resonances~\cite{chernyak2020bound,zakharov2020transverse}, while tailoring the resonant mode structure provides control over the spectral position and magnitude of the magneto-optical response~\cite{mamian2025tailoring}.

For optomagnetic applications, Mie-resonant garnet particles are especially interesting because the optical near field can generate spatially inhomogeneous effective magnetic fields through the inverse Faraday effect (IFE)~\cite{krichevsky2024inverse}. The magnitude and symmetry of this effective field are determined by the local distribution of the optical field inside the magnetic material. As a result, different Mie modes can produce different optomagnetic-field profiles and may be used to address magnetic excitations selectively. A related principle was demonstrated in magnetophotonic structures, where a spatially inhomogeneous inverse Faraday effect enabled selective excitation of exchange-dominated spin waves~\cite{krichevsky2024spatially}. These results make the localization of the resonant optical field inside the garnet volume a key design parameter for nanoscale optomagnonics.

However, implementing efficient Mie resonances in epitaxial garnet nanostructures is not straightforward. High-quality garnet films are commonly grown on GGG substrates because of their compatibility with iron-garnet materials. At the same time, the refractive-index contrast between a garnet nanodisk and a homogeneous GGG substrate is relatively small. As a consequence, the optical mode of a nanodisk on a homogeneous GGG substrate is not fully confined inside the disk: part of the resonant field penetrates into the substrate, reducing the field enhancement inside the magnetic volume. Replacing GGG with a substrate of lower refractive index could improve optical confinement, but this would compromise the main materials advantage of the system, namely compatibility with the growth of high-quality garnet films. Alternative integration routes on non-garnet platforms often require additional crystallization, seed-layer, or local thermal-processing strategies~\cite{sun2015single, hayashi2024magneto}.

Recent works have explored two-sided Bragg-cavity geometries, where a resonant element is embedded between upper and lower Bragg reflectors. In particular, dielectric micropillar resonators sandwiched between Bragg mirrors were proposed to enhance the quality-factor-to-mode-volume ratio through a quasi-BIC regime~\cite{kolodny2019q}. A related concept was recently extended to magneto-optical metasurfaces, where Mie resonators and garnet films were introduced as defect layers inside a one-dimensional photonic-crystal cavity to obtain high-Q resonances and enhanced Faraday rotation~\cite{yang2025giant}. These approaches rely on two-sided Bragg confinement of the resonant element. In contrast, here we focus on a simpler one-sided geometry that can be implemented within established all-garnet epitaxial growth technologies. All-garnet magneto-optical photonic crystals based on alternating magnetic and nonmagnetic garnet layers have been experimentally realized on garnet substrates using pulsed-laser deposition and magnetron sputtering~\cite{kahl2004enhanced,khartsev2007high}. This makes it possible to grow the photonic multilayer together with the magnetic top layer on a flat substrate and subsequently pattern only the upper BIG layer into a nanoresonator.

In this work, we consider a single epitaxial BIG nanodisk placed on a one-dimensional SGG/BIG photonic-crystal multilayer grown on GGG. The photonic band gap of the multilayer is chosen to overlap with the spectral region of the nanodisk Mie resonances. One-dimensional garnet magnetophotonic crystals are known to support photonic band gaps and to enhance magneto-optical effects through interference and resonant light localization in multilayer structures~\cite{berzhansky2011one}. Unlike hybrid garnet/SiO\textsubscript{2} photonic-crystal designs, the present SGG/BIG multilayer is composed entirely of garnet materials and therefore remains compatible with epitaxial growth on GGG. The resulting one-sided structure preserves direct top-side optical access to the nanodisk while the underlying photonic crystal suppresses downward leakage of the resonant field and increases its localization inside the BIG disk. We show that this modification of the optical environment enhances the local electromagnetic field and the optomagnetic response and, at the same time, modifies the spectral positions and relative strengths of the electric- and magnetic-dipole contributions, as well as the spatial distribution of the optical field inside the magnetic material. These results demonstrate that the resonant response of an individual magnetic nanodisk can be substantially modified through its photonic environment, enabling wavelength-dependent control of the magnitude and spatial distribution of the local IFE-related response without changing the geometry of the nanodisk.

\section{Model and numerical methods}
\subsection{Structure under study}

We consider two configurations of a single BIG nanoresonator, as shown in Fig.~\ref{fig:structure_scheme}. The reference structure consists of a bismuth-substituted iron garnet, \(\mathrm{Bi_{2.97}Er_{0.03}Fe_{4.00}Al_{0.5}Ga_{0.5}O_{12}}\) (BIG) nanodisk placed directly on a homogeneous gadolinium gallium garnet, \(\mathrm{Gd_3Ga_5O_{12}}\) (GGG), substrate. In the modified configuration, the same nanodisk is placed on a one-dimensional photonic-crystal multilayer grown on GGG. In both cases, the disk radius and height are fixed at $r = 175~\mathrm{nm}, \: h = 225~\mathrm{nm}$. Using identical disk parameters allows us to isolate the role of the substrate in the optical response of the resonator.

The photonic crystal substrate consists of ten pairs of BIG and samarium gallium garnet, \(\mathrm{Sm_3Ga_5O_{12}}\) (SGG), layers. Starting from the GGG substrate side, the multilayer sequence is written as $[\mathrm{BIG}/\mathrm{SGG}]^{10},$
so that the uppermost layer adjacent to the nanodisk is SGG. The layer thicknesses were chosen according to the quarter-wave condition $d_i = \frac{\lambda_0}{4 n_i(\lambda_0)},$
where $d_i$ and $n_i$ are the thickness and refractive index of the $i$-th layer, respectively. The central wavelength of the photonic band gap was set to \(\lambda_0 = 800~\mathrm{nm}\), which gives
\[
d_{\mathrm{BIG}} = 78~\mathrm{nm}, \qquad
d_{\mathrm{SGG}} = 102~\mathrm{nm}.
\]
The total thickness of the ten-period photonic-crystal multilayer is therefore \(1.8~\mu\mathrm{m}\).

\begin{IFEcolumnfigure}
    \centering
   \includegraphics[width=\linewidth]{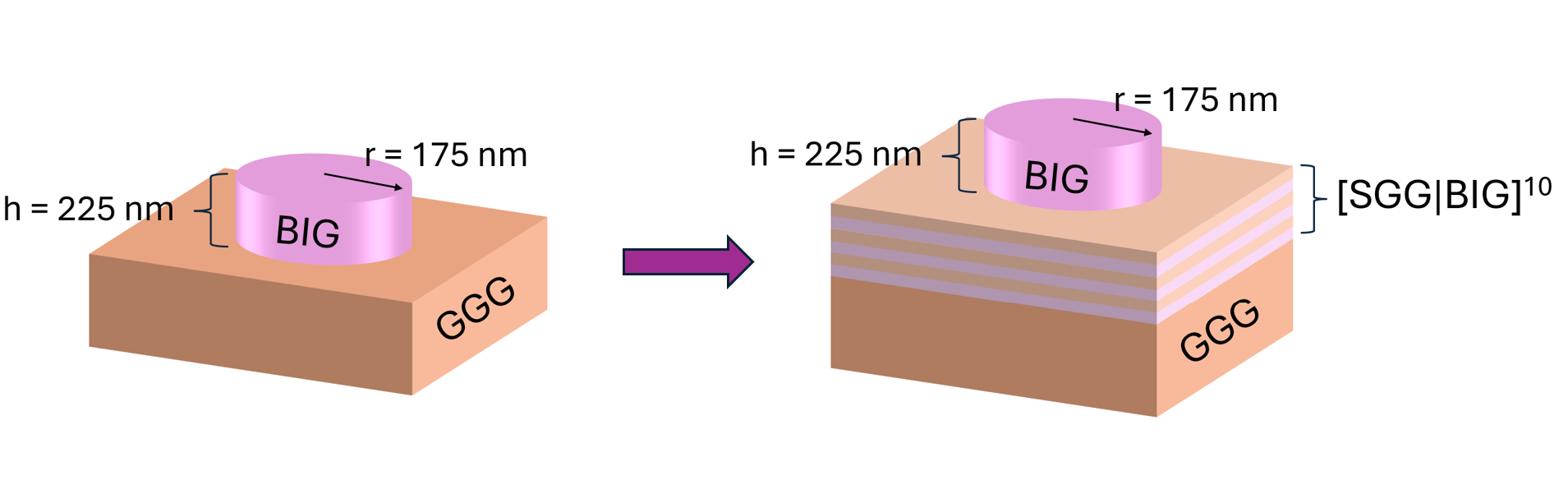}
    \caption{Scheme of the structures under study: a BIG nanodisk on a homogeneous GGG substrate and a BIG nanodisk on a one-dimensional SGG/BIG photonic crystal multilayer grown on GGG. The nanodisk radius and height are \(r=175~\mathrm{nm}\) and \(h=225~\mathrm{nm}\), respectively. The multilayer consists of ten quarter-wave SGG/BIG pairs with the photonic band-gap center at \(\lambda_0=800~\mathrm{nm}\).}
    \label{fig:structure_scheme}
\end{IFEcolumnfigure}

The wavelength-dependent optical constants of BIG, SGG, and GGG were taken from~\cite{ignatyeva2024asymmetric}; the corresponding dispersion curves are shown in Appendix~A. All-garnet BIG/SGG magneto-optical photonic crystals grown on garnet substrates have been experimentally realized and structurally characterized by X-ray diffraction, including structures designed for operation near 750~nm~\cite{khartsev2007high}. Related BIG/SGG photonic-crystal structures have also been demonstrated at a resonance wavelength of 980~nm~\cite{dzibrou2009fitting}. Heteroepitaxial SGG/substituted-BIG photonic crystals were fabricated on GGG(111) by radio-frequency magnetron sputtering~\cite{khartsev2011high}. These results support the feasibility of using heteroepitaxial BIG/SGG multilayers on garnet substrates as the photonic platform considered here.

\subsection{Numerical methods}

Numerical simulations were performed using the finite-difference time-domain (FDTD) method. The optical fields were calculated using the wavelength-dependent scalar dielectric permittivities of BIG, SGG, and GGG given in Appendix~A. These fields were subsequently used to evaluate the IFE-related response. Perfectly matched layer (PML) boundary conditions were applied in all directions. 
Scattering spectra, multipolar contributions, and $|E|^2$ maps were calculated for normally incident linearly polarized excitation, with the electric field polarized along $x$ and the wave propagating along $-z$. All IFE-related spectra and maps were calculated for incident circularly polarized radiation. Circular polarization was generated using two linearly polarized plane-wave sources ($E_x$ and $E_y$) of equal amplitude with a $90^\circ$ phase offset. All electric-field intensity and IFE-related maps are normalized by $|E_0|^2=|E_{0x}|^2+|E_{0y}|^2$, where $E_{0x}$ and $E_{0y}$ are the complex amplitudes of the incident electric-field components. At normal incidence along the disk axis, the incident wave has $E_{0z}=0$. The quantity $|E_0|^2$ is proportional to the total incident intensity and includes both orthogonal components for circular polarization.

The ED, MD, EQ, and MQ moments are evaluated using volume integrals of the induced electric current density over the BIG nanodisk~\cite{alaee2018electromagnetic}. The fields recorded within this volume are calculated for the complete structure and therefore include the influence of the substrate. The multipolar scattering contributions are evaluated using free-space radiation formulas for the nanodisk in air, on homogeneous GGG, and on the SGG/BIG photonic-crystal substrate; their plotted sum contains only the ED, MD, EQ, and MQ contributions of the nanodisk. The individual contributions and their sum characterize the induced multipolar response of the nanodisk in the different optical environments. The electric-field and IFE-related enhancements are evaluated directly from the simulated local fields.

\section{Results and discussion}
\subsection{Optical response and modal composition}

Figure~\ref{fig:optical_response} compares the scattering response and multipolar decomposition for three configurations: an isolated BIG nanodisk in air, the same nanodisk on homogeneous GGG, and the nanodisk on the BIG/SGG photonic crystal substrate. For the isolated nanodisk in air, Fig.~\ref{fig:optical_response}(a), the response is predominantly dipolar, with a broad electric-dipole contribution strongly overlapping a narrower magnetic-dipole peak.

For the nanodisk on homogeneous GGG, the multipolar sum exhibits a single broad maximum, Fig.~\ref{fig:optical_response}(b). We use $\lambda=800$~nm, near this maximum, as the reference wavelength for the spatial maps. The response is dominated by the electric-dipole contribution, while the magnetic-dipole contribution remains substantially weaker over the considered spectral range.
\end{multicols}
\begin{IFEplacedfigure}
    \centering

    \begin{minipage}[t]{0.31\textwidth}
        \centering
        \includegraphics[width=\linewidth]{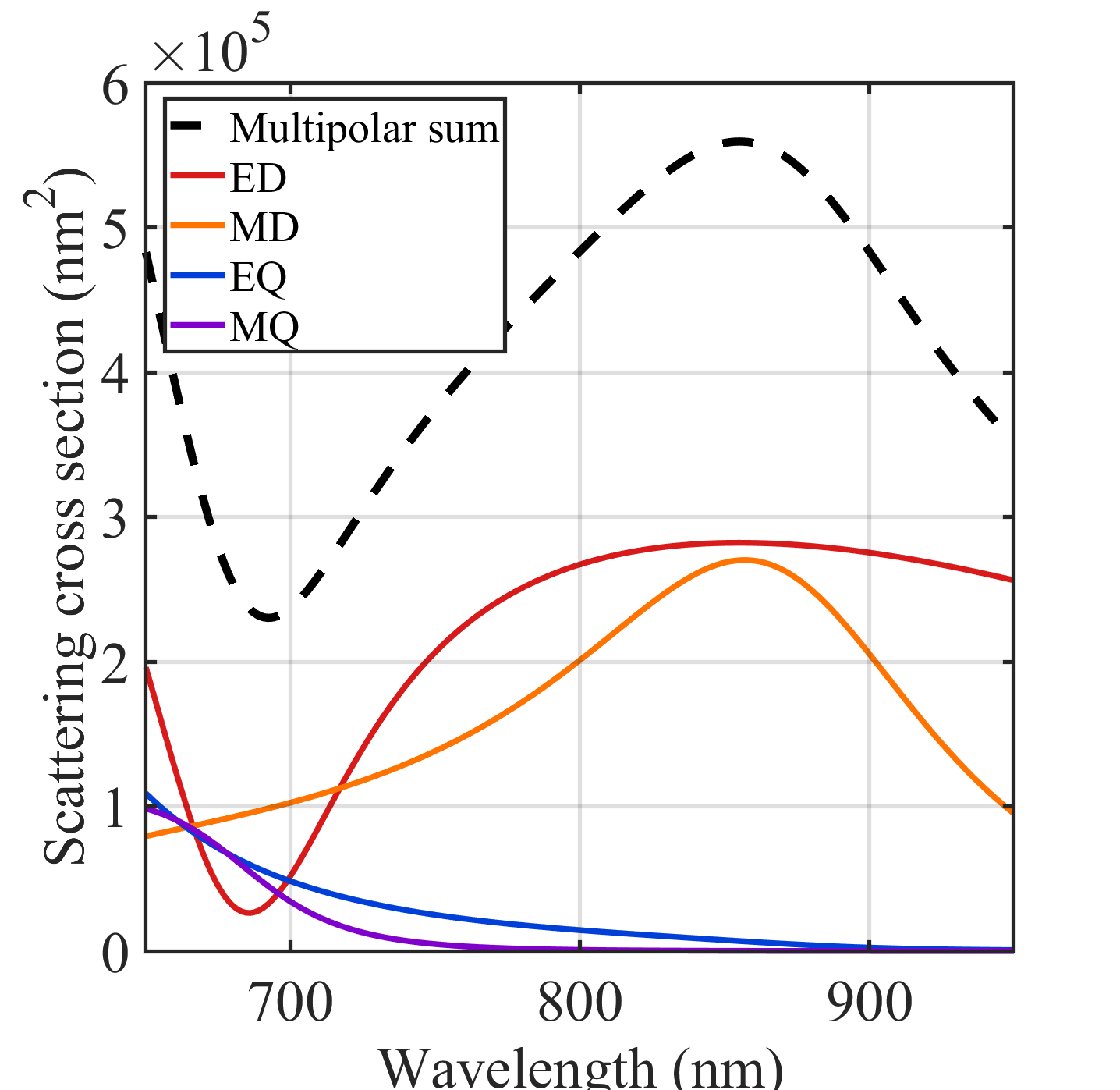}
        \textbf{(a)}
    \end{minipage}
    \hfill
    \begin{minipage}[t]{0.31\textwidth}
        \centering
        \includegraphics[width=\linewidth]{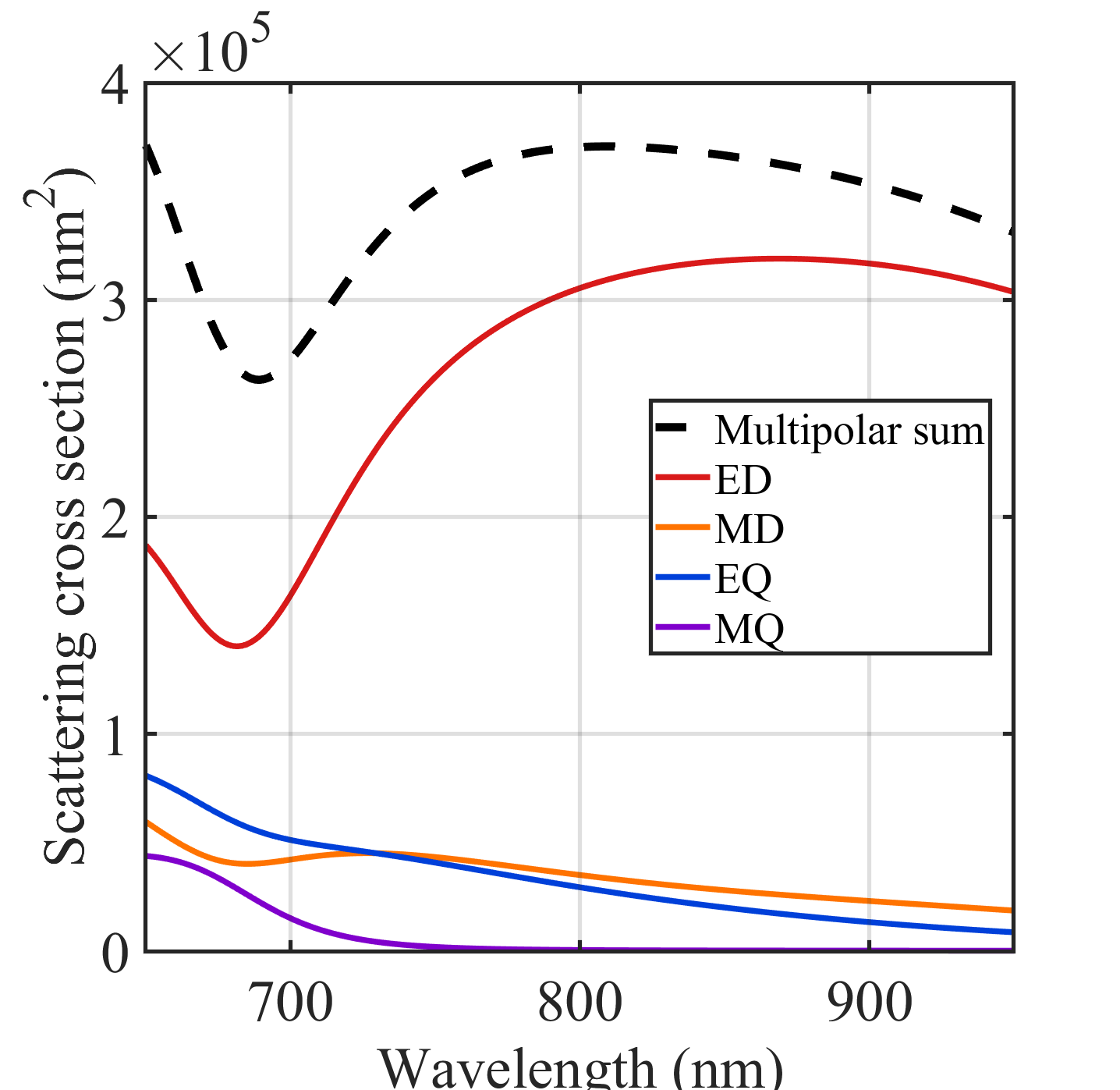}
        \textbf{(b)}
    \end{minipage}
    \hfill
    \begin{minipage}[t]{0.31\textwidth}
        \centering
        \includegraphics[width=\linewidth]{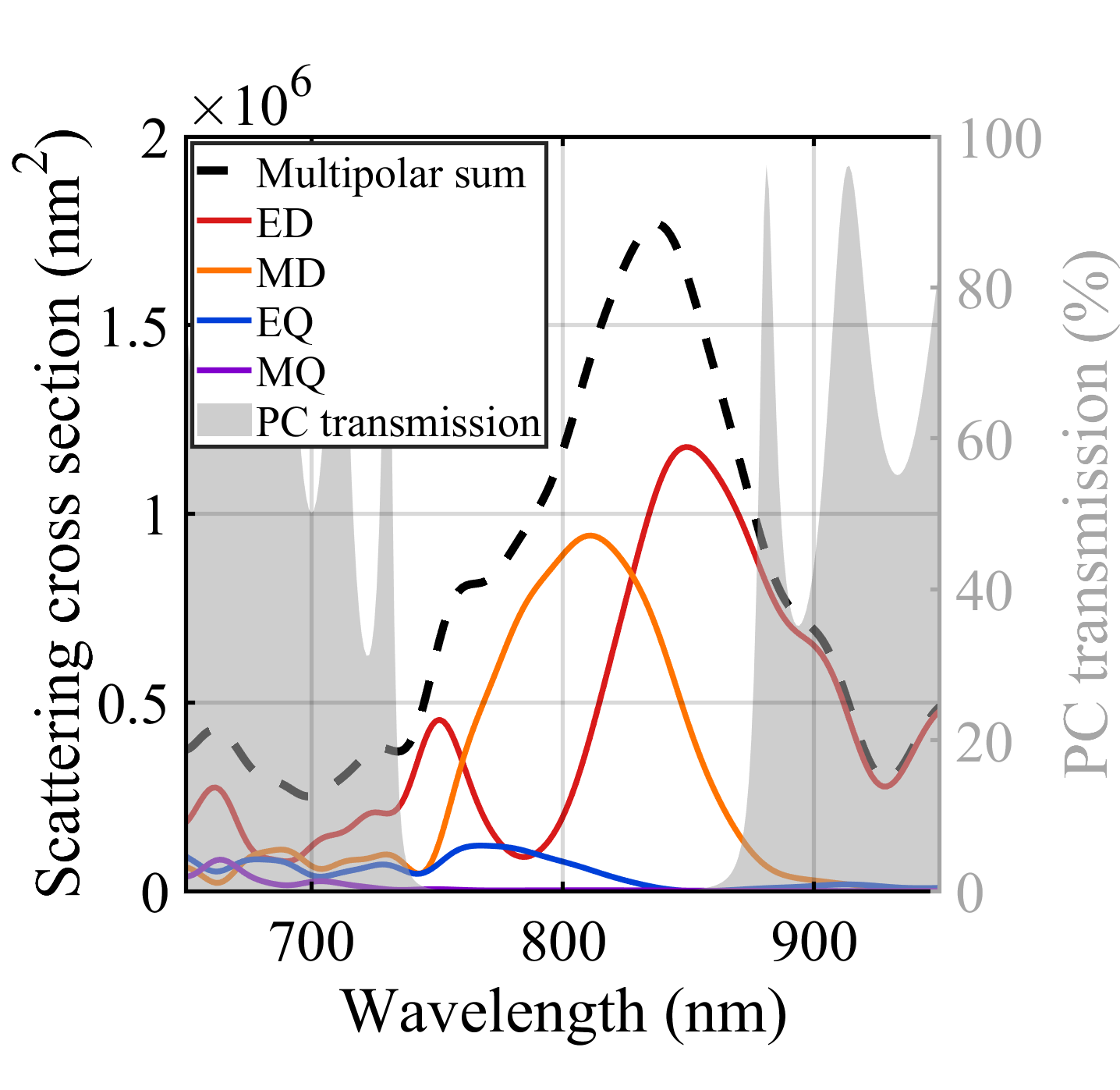}
        \textbf{(c)}
    \end{minipage}

    \caption{
    Scattering response and multipolar decomposition of the BIG nanodisk for
    (a) an isolated disk in air,
    (b) a disk on homogeneous GGG,
    and (c) a disk on the SGG/BIG photonic crystal substrate.
    The colored curves show the electric-dipole (ED), magnetic-dipole (MD), electric-quadrupole (EQ), and magnetic-quadrupole (MQ) scattering contributions calculated from the multipole moments induced in the nanodisk. The black dashed curves show the sum of these four contributions.
    }
    \label{fig:optical_response}
\end{IFEplacedfigure}
\begin{multicols}{2}
A qualitatively different behavior is obtained for the nanodisk on the SGG/BIG photonic-crystal substrate. In Fig.~\ref{fig:optical_response}(c), the gray shaded area shows the normal-incidence transmission spectrum of the bare photonic crystal; the low-transmission region corresponds to the photonic bandgap of the multilayer. The multilayer was designed with the normal-incidence photonic bandgap centered near $\lambda_0 = 800~\mathrm{nm}$. Compared with the nanodisk on homogeneous GGG, the nanodisk on the photonic crystal exhibits more pronounced electric- and magnetic-dipole contributions. The corresponding resonances lie close to each other and substantially overlap, producing a single broad maximum in the summed multipolar response in this spectral region. The sum of the ED, MD, EQ, and MQ scattering contributions reaches a pronounced maximum near $\lambda=837$~nm, where both electric- and magnetic-dipole contributions are substantial. The MD and ED contributions reach their respective maxima near $\lambda=811$~nm and $849$~nm. At the MD maximum, the magnetic-dipole contribution exceeds the electric-dipole contribution by a factor of approximately 2.3. At the ED maximum, the electric-dipole contribution exceeds the magnetic-dipole contribution by a factor of approximately 2.5. These wavelengths therefore correspond to MD- and ED-dominated regimes, respectively. Thus, the photonic-crystal substrate modifies both the spectral positions and relative weights of the induced dipolar contributions.

We next examine how the modification of the nanodisk optical response by the photonic-crystal substrate affects the inverse Faraday effect inside the nanodisk.

\subsection{Optomagnetic response induced by the inverse Faraday effect}

The inverse Faraday effect (IFE) is a nonlinear magneto-optical phenomenon in which circularly polarized light induces a static or transient magnetization in a medium. This effect was first described within a thermodynamic approach by Pitaevskii~\cite{pitaevskii1961electric} and was later analyzed in terms of nonlinear optical response and Raman-type light--matter interaction~\cite{pershan1966theoretical}. In magnetic dielectrics, the IFE provides a mechanism for optical excitation of magnetization dynamics without applying an external magnetic-field pulse. In particular, circularly polarized femtosecond pulses can act as ultrashort effective magnetic-field pulses and excite coherent spin precession~\cite{kimel2005ultrafast,hansteen2006nonthermal}. This mechanism is also closely related to helicity-dependent all-optical magnetic recording, where the magnetic response is controlled by the polarization state of the excitation pulse~\cite{stanciu2007all,stupakiewicz2019selection}.

In the classical thermodynamic description proposed by Pitaevskii, circularly polarized light in a transparent medium induces an effective magnetic field associated with the inverse Faraday effect. This field can be written in Gaussian units as
\begin{equation}
\mathbf{H}^{\mathrm{IFE}}
=
-\frac{i g}{16\pi M}
\left[
\mathbf{E}(\omega)\times \mathbf{E}^{*}(\omega)
\right],
\label{eq:pitaevskii}
\end{equation}
where $g$ is the magneto-optical gyration coefficient, $M$ is the magnitude of the magnetization of the medium, $\mathbf{E}(\omega)$ is the complex electric-field amplitude at the optical frequency, and $\mathbf{E}^{*}(\omega)$ is its complex conjugate~\cite{pitaevskii1961electric}.

Since the vector $\mathbf{E}(\omega)\times\mathbf{E}^{*}(\omega)$ is purely imaginary, Eq.~\eqref{eq:pitaevskii} is proportional to the real axial vector $\mathrm{Im}[\mathbf{E}(\omega)\times\mathbf{E}^{*}(\omega)]$. In the present work, we do not calculate the absolute value of the effective magnetic field, because this would require the composition-dependent coefficient $g/M$ for BIG. Therefore, we use the field-dependent part of Eq.~\eqref{eq:pitaevskii} as a relative measure of the local optomagnetic response and define
\begin{equation}
|\mathrm{IFE}|=\left|\mathrm{Im}\left[\mathbf{E}(\omega)\times\mathbf{E}^{*}(\omega)\right]\right|,
\label{eq:ife_abs}
\end{equation}
where $|\mathrm{IFE}|$ is proportional to the magnitude of the local IFE-related response up to a composition-dependent coefficient.

Here $\mathrm{IFE}_x$, $\mathrm{IFE}_y$, and $\mathrm{IFE}_z$ denote the Cartesian components of the optical factor $\mathrm{Im}[\mathbf{E}\times\mathbf{E}^{*}]$.

Unlike the electric-field intensity $|E|^2$, the quantity $|\mathrm{IFE}|$ depends on the complex vector structure of the optical field. It vanishes for a purely linearly polarized field and becomes nonzero when the local field has an elliptical or circular polarization component. Therefore, the $|\mathrm{IFE}|$ maps characterize the part of the resonant near field that can contribute to the inverse Faraday effect, rather than only the spatial distribution of field intensity.

For magnetophotonic structures, resonant optical confinement can enhance the optomagnetic response by increasing the optical field inside the magnetic material. This principle has been demonstrated in magnetophotonic microcavities, where localization of the optical field inside a magnetic layer leads to a strong enhancement and subwavelength localization of the IFE-induced effective field within the photonic bandgap~\cite{kozhaev2018giant}.

In the present work, we apply this approach to a BIG nanodisk placed either on a homogeneous GGG substrate or on an SGG/BIG photonic crystal substrate. Since the same BIG material is used in all considered configurations, comparison of $|\mathrm{Im}[\mathbf{E}\times\mathbf{E}^{*}]|$ directly shows how the photonic environment modifies the field-dependent part of the optomagnetic response inside the magnetic nanodisk.

\end{multicols}
\begin{IFEplacedfigure}
\centering
\includegraphics[width=0.92\textwidth]{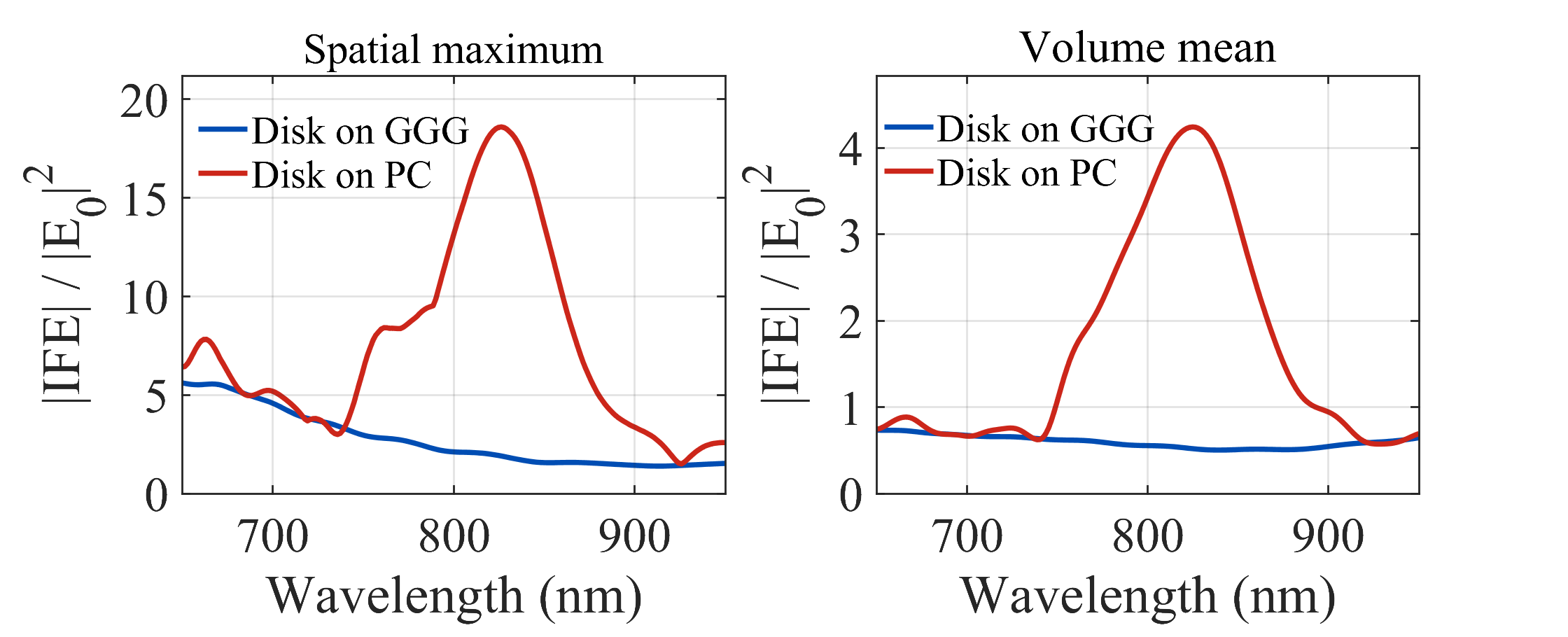}
\par\smallskip
\caption{Spectral dependence of the spatial maximum $Q_{\mathrm{max}}$ (left) and the volume average $Q_{\mathrm{mean}}$ (right) of the normalized IFE-related quantity $|\mathrm{Im}[\mathbf{E}\times\mathbf{E}^{*}]|/|E_0|^2$ inside the BIG nanodisk under normally incident circularly polarized radiation. The blue and red curves correspond to the nanodisk on homogeneous GGG and on the SGG/BIG photonic-crystal substrate, respectively. Both quantities are evaluated over the entire nanodisk volume according to Eq.~\eqref{eq:IFE_spectral_metrics}.}
\label{fig:IFE_spectra}
\end{IFEplacedfigure}
\begin{multicols}{2}

Figure~\ref{fig:IFE_spectra} compares the spectral dependence of the spatial maximum and the volume average of the normalized IFE-related quantity for the BIG nanodisk on homogeneous GGG and on the SGG/BIG photonic-crystal substrate. At each wavelength, these quantities are calculated over the entire nanodisk volume as
\begin{equation}
\begin{aligned} Q_{\mathrm{max}}(\lambda)&=\max_{\mathbf{r}\in V_{\mathrm{d}}}\frac{|\mathrm{IFE}(\mathbf{r},\lambda)|}{|E_0|^2},\\ Q_{\mathrm{mean}}(\lambda)&=\frac{1}{V_{\mathrm{d}}}\int_{V_{\mathrm{d}}}\frac{|\mathrm{IFE}(\mathbf{r},\lambda)|}{|E_0|^2}\,\mathrm{d}^{3}\mathbf{r}, \end{aligned}
\label{eq:IFE_spectral_metrics}
\end{equation}
where $V_{\mathrm{d}}$ is the volume of the BIG nanodisk, $\mathbf{r}$ denotes position within the disk, and $|E_0|^2$ is the squared magnitude of the total incident electric-field amplitude. Thus, $Q_{\mathrm{max}}$ characterizes the strongest local response at each wavelength, while $Q_{\mathrm{mean}}$ describes the magnitude of the response averaged over the magnetic volume.

Both spectra exhibit a broad enhancement band in the region where the ED and MD resonances overlap, close to the maximum of the summed multipolar scattering contributions at $\lambda=837$~nm. In this band, the response of the nanodisk on the photonic crystal substantially exceeds that of the same nanodisk on homogeneous GGG.

For each of the two spectra, we take the peak value for the disk on the photonic crystal and divide it by the value for the disk on GGG at the same wavelength. The resulting enhancement factors are approximately $9.8$ for the spatial maximum and $8.2$ for the volume average.

To examine the spatial localization of the optical field, Fig.~\ref{fig:E2_maps} compares the normalized electric-field intensity for the reference nanodisk on homogeneous GGG and the nanodisk on the SGG/BIG photonic-crystal substrate. The upper row, Fig.~\ref{fig:E2_maps}(a,b), shows horizontal $xy$-cross-sections, while the lower row, Fig.~\ref{fig:E2_maps}(c,d), shows vertical $yz$-cross-sections near the disk center. For each structure and wavelength, the horizontal $xy$-cross-section is taken at the height where $|E|^2$ reaches its maximum inside the BIG nanodisk. This allows us to compare the strongest field localization inside the magnetic volume, while the vertical cross-sections show how the resonant field is distributed along the disk height and into the substrate.

For the nanodisk on homogeneous GGG, the electric field is only weakly localized inside the disk. In Fig.~\ref{fig:E2_maps}(a), the maximum value of $|E|^2/|E_0|^2$ inside the BIG nanodisk is $2.2$. The vertical section in Fig.~\ref{fig:E2_maps}(c) shows that the resonant field extends noticeably into the homogeneous substrate. This agrees with the broad and weakly structured optical response in Fig.~\ref{fig:optical_response}(b): the GGG substrate provides an efficient leakage channel for the disk mode.

At $\lambda=837$~nm, the maximum normalized electric-field intensity $|E|^2/|E_0|^2$ within the plotted disk cross-section reaches approximately 17.5, corresponding to an enhancement factor of 7.8 relative to the reference nanodisk on GGG at $800$~nm. The horizontal map in Fig.~\ref{fig:E2_maps}(b) shows pronounced field concentration inside the nanodisk, while the vertical section in Fig.~\ref{fig:E2_maps}(d) demonstrates that the field rapidly decays below the disk--photonic crystal interface. The enhanced field intensity inside the nanodisk, together with its decay into the multilayer, is consistent with improved optical confinement provided by the photonic-crystal substrate.

Strong field localization is also obtained at $811$ and $849$~nm, within the same band of overlapping dipolar resonances; the corresponding maps are presented in Appendix~B (Fig.~\ref{fig:E2_maps_appendix}). Among the three examined wavelengths, the largest electric-field enhancement is obtained at $837$~nm, at the maximum of the summed multipolar scattering contributions.

\end{multicols}
\begin{IFEplacedfigure}
\centering
\begin{minipage}[b]{0.43\textwidth}
\centering
\includegraphics[width=\linewidth,height=50mm,keepaspectratio,trim={11.29927pt 33.12164pt 21.82327pt 49.68246pt},clip]{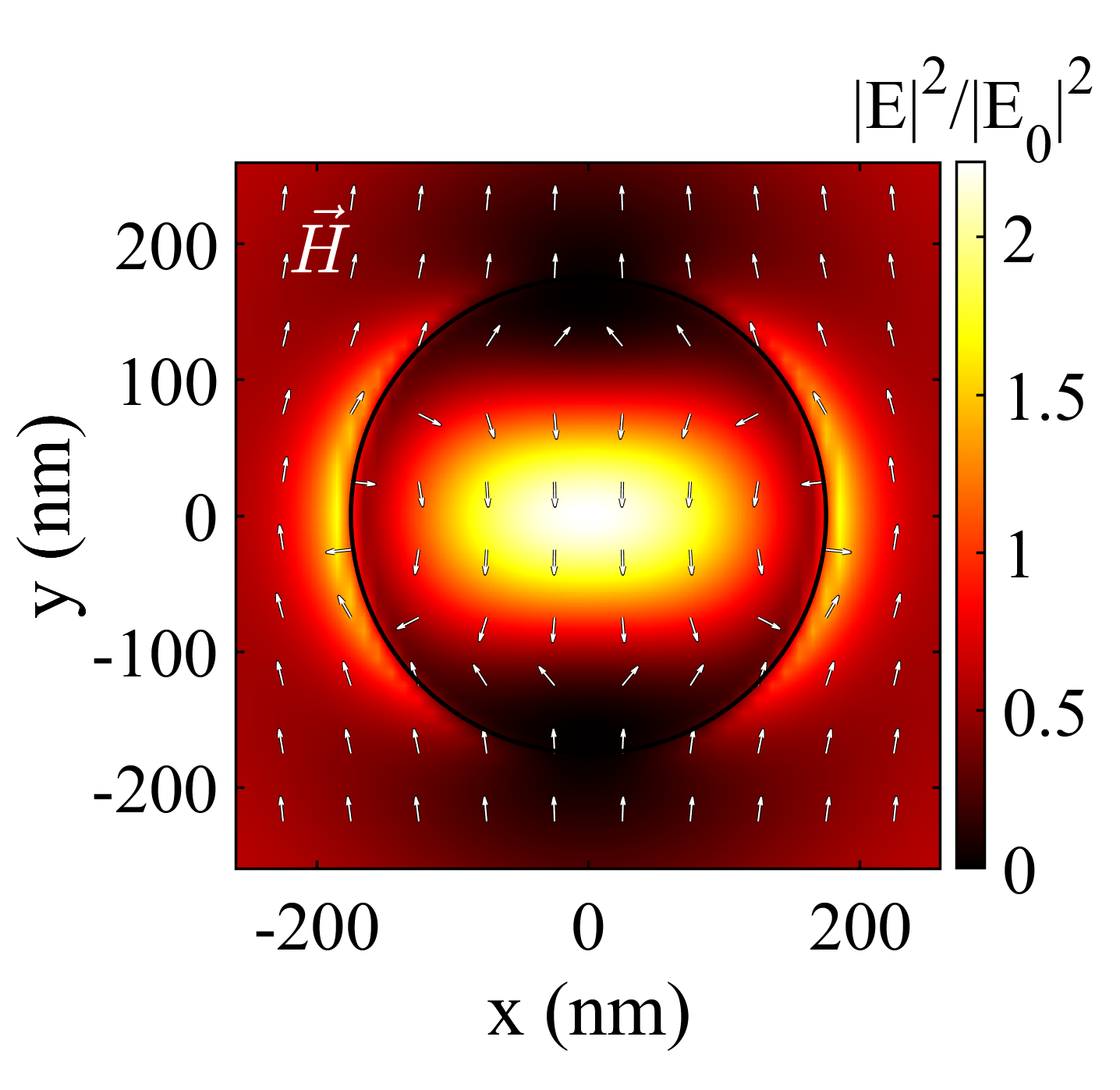}
\par\vspace{1.5mm}
(a)
\end{minipage}%
\hspace{0.025\textwidth}%
\begin{minipage}[b]{0.43\textwidth}
\centering
\includegraphics[width=\linewidth,height=50mm,keepaspectratio,trim={10.5405pt 30.85896pt 36.1412pt 27.1039pt},clip]{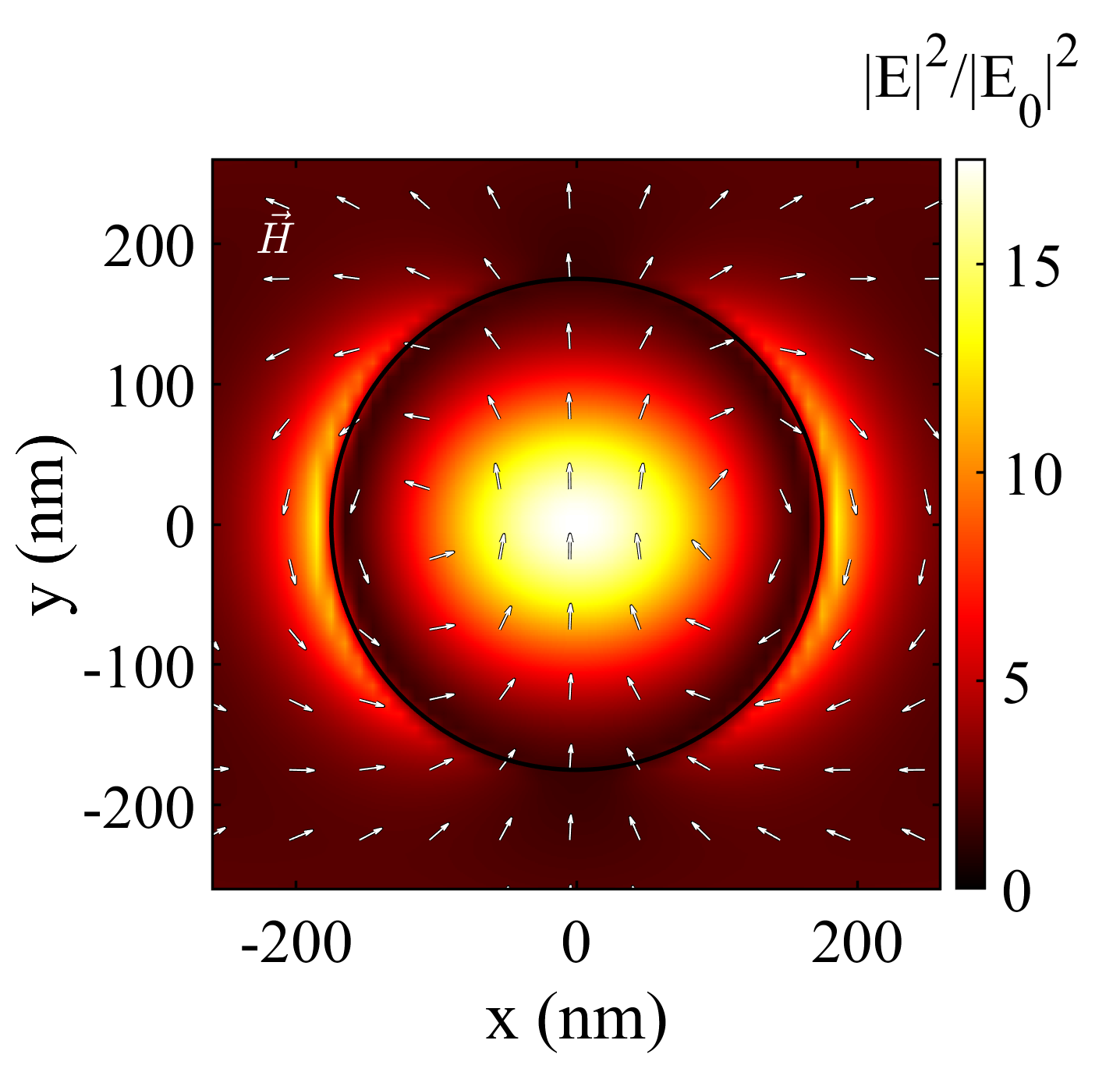}
\par\vspace{1.5mm}
(b)
\end{minipage}
\par\vspace{3mm}
\begin{minipage}[b]{0.43\textwidth}
\centering
\includegraphics[width=\linewidth,height=50mm,keepaspectratio,trim={11.29927pt 0.0pt 1.50107pt 0.0pt},clip]{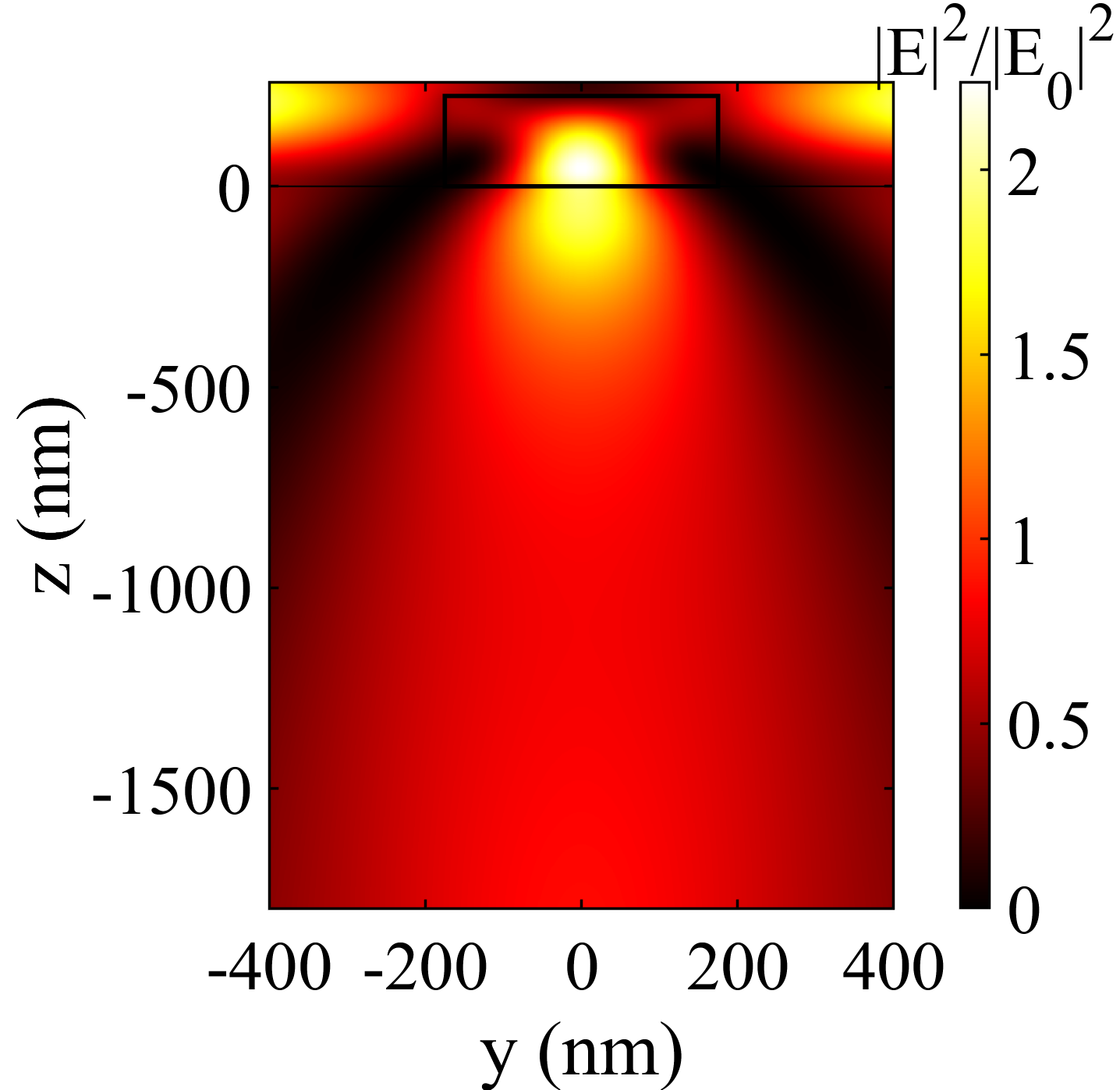}
\par\vspace{1.5mm}
(c)
\end{minipage}%
\hspace{0.025\textwidth}%
\begin{minipage}[b]{0.43\textwidth}
\centering
\includegraphics[width=\linewidth,height=50mm,keepaspectratio,trim={10.5405pt 0.0pt 26.34299pt 0.0pt},clip]{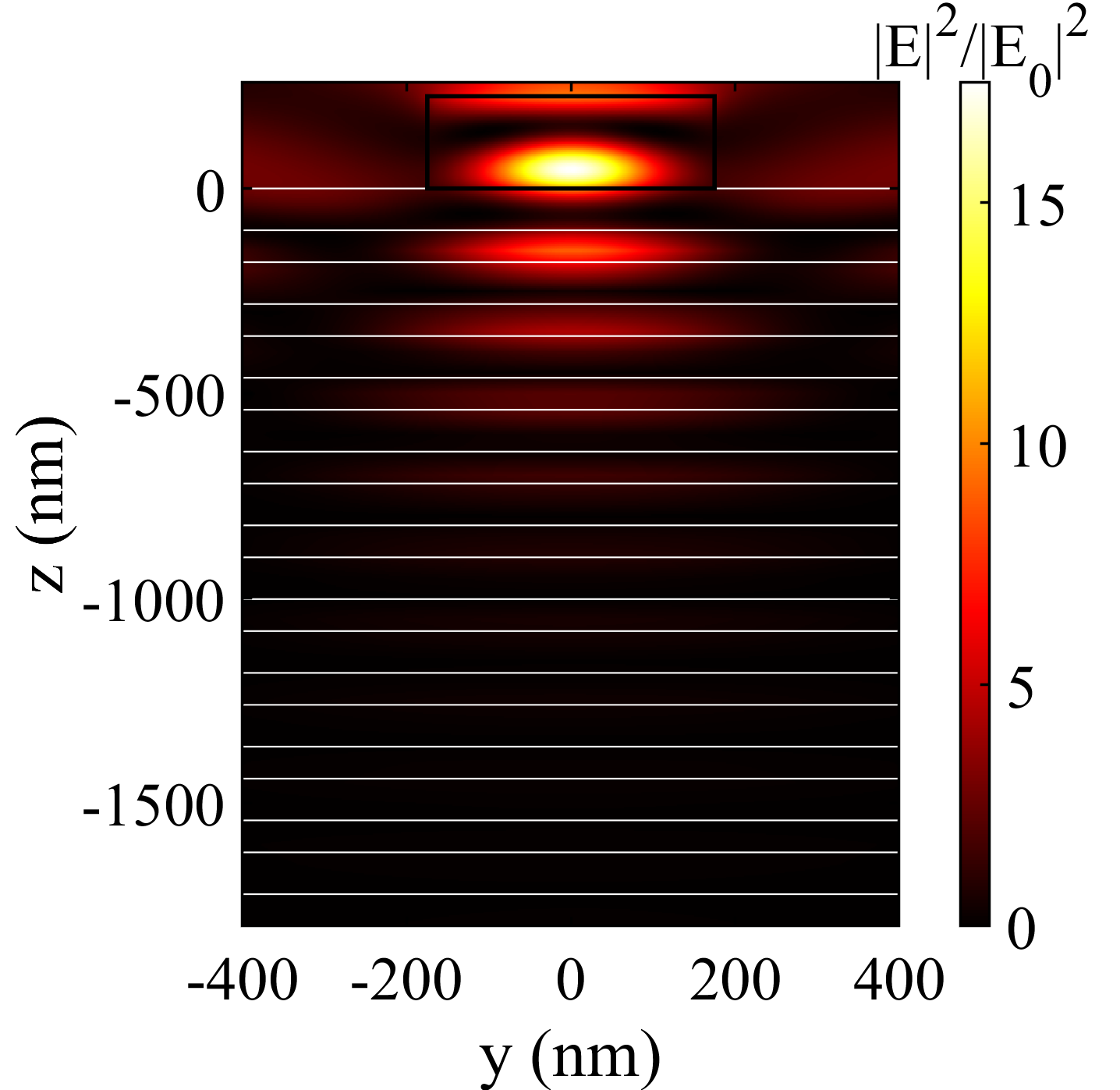}
\par\vspace{1.5mm}
(d)
\end{minipage}
\par\smallskip
\caption{Near-field distributions of the normalized electric-field intensity, $|E|^2/|E_0|^2$, for the BIG nanodisk under linearly polarized radiation, where $E_0$ is the amplitude of the incident electric field. The upper row shows horizontal $xy$-cross-sections taken at the height of maximum $|E|^2$ inside the nanodisk, while the lower row shows vertical $yz$-cross-sections near the disk center. Panels (a,c) correspond to the nanodisk on homogeneous GGG at $\lambda=800$~nm, and panels (b,d) to the nanodisk on the SGG/BIG photonic-crystal substrate at $\lambda=837$~nm. The wavelength $\lambda=837$~nm corresponds to the maximum of the summed ED, MD, EQ, and MQ scattering contributions for the nanodisk on the photonic crystal. The color scales show the calculated values of $|E|^2/|E_0|^2$. The black contour marks the boundary of the BIG nanodisk, and the white horizontal lines indicate the interfaces of the SGG/BIG photonic-crystal layers. In the horizontal cross-sections, the white arrows show the instantaneous in-plane optical magnetic field, $(H_x,H_y)$, at a fixed optical phase.}
\label{fig:E2_maps}
\end{IFEplacedfigure}
\begin{multicols}{2}

\subsubsection{Out-of-plane IFE component for in-plane magnetization}

The relevance of each IFE component to magnetic excitation depends on its orientation relative to the equilibrium magnetization. For a uniform magnetization in the disk plane, the axial effective field is transverse to the magnetization and can drive precession. We therefore examine $|\mathrm{IFE}_z|$ separately. In the examined cross-sections, this component has the largest peak magnitude. A strong axial IFE response at the electric-dipole resonance was discussed in~\cite{krichevsky2024inverse}; in the present structure, both dipolar contributions coexist within the main resonance band. The peak value of a multipolar scattering contribution alone does not determine the local IFE magnitude, which depends on the amplitudes and relative phases of the local electric-field components.

\end{multicols}
\begin{IFEplacedfigure}
\centering
\begin{minipage}[b]{0.43\textwidth}
\centering
\includegraphics[width=\linewidth,height=50mm,keepaspectratio,trim={10.48053pt 60.52994pt 115.27266pt 64.31458pt},clip]{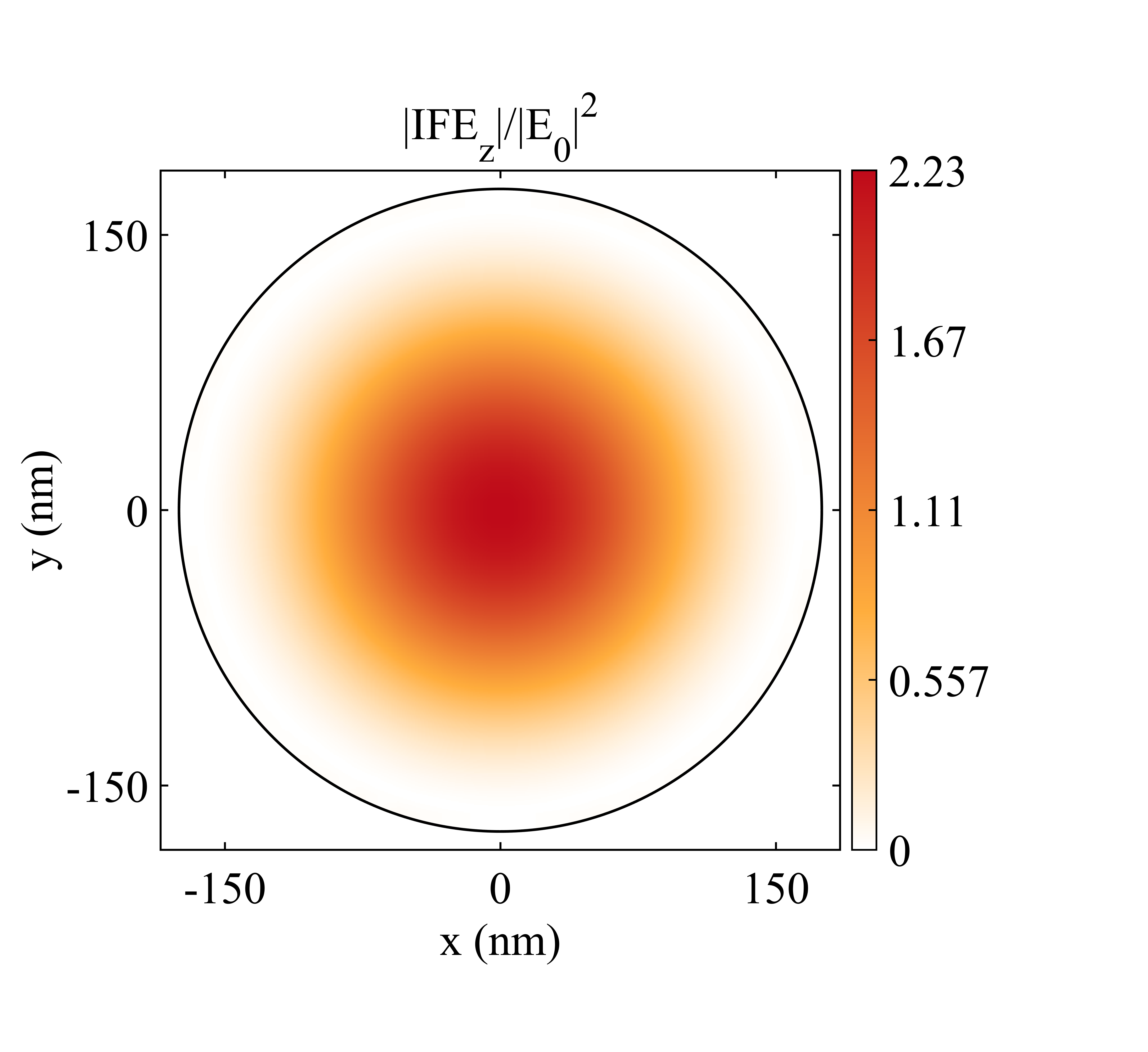}
\par\vspace{1.5mm}
(a)
\end{minipage}%
\hspace{0.025\textwidth}%
\begin{minipage}[b]{0.43\textwidth}
\centering
\includegraphics[width=\linewidth,height=50mm,keepaspectratio,trim={10.48053pt 60.52994pt 132.43585pt 64.31458pt},clip]{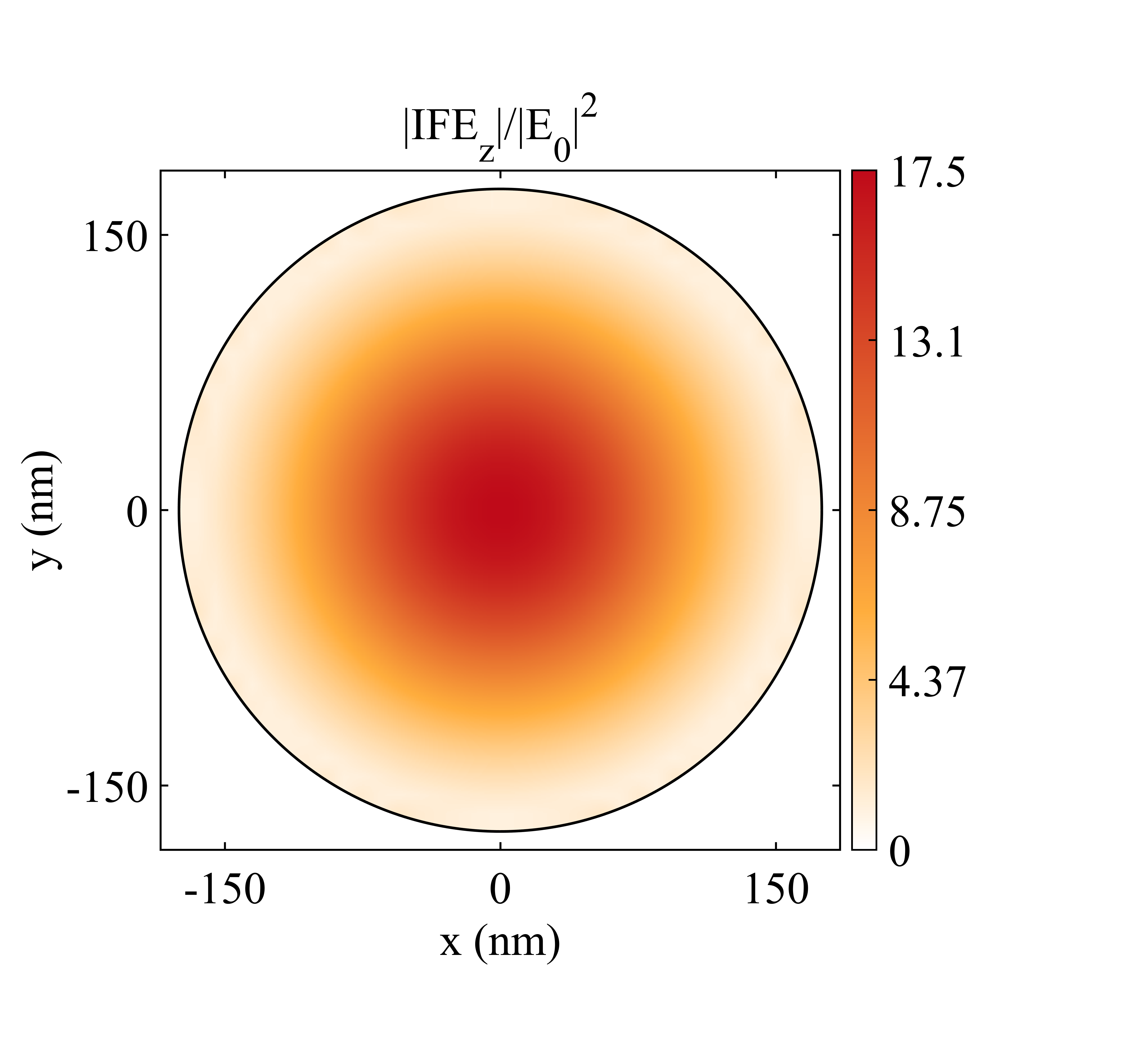}
\par\vspace{1.5mm}
(b)
\end{minipage}
\par\vspace{3mm}
\begin{minipage}[b]{0.43\textwidth}
\centering
\includegraphics[width=\linewidth,height=50mm,keepaspectratio,trim={22.40346pt 52.39716pt 112.55989pt 61.06754pt},clip]{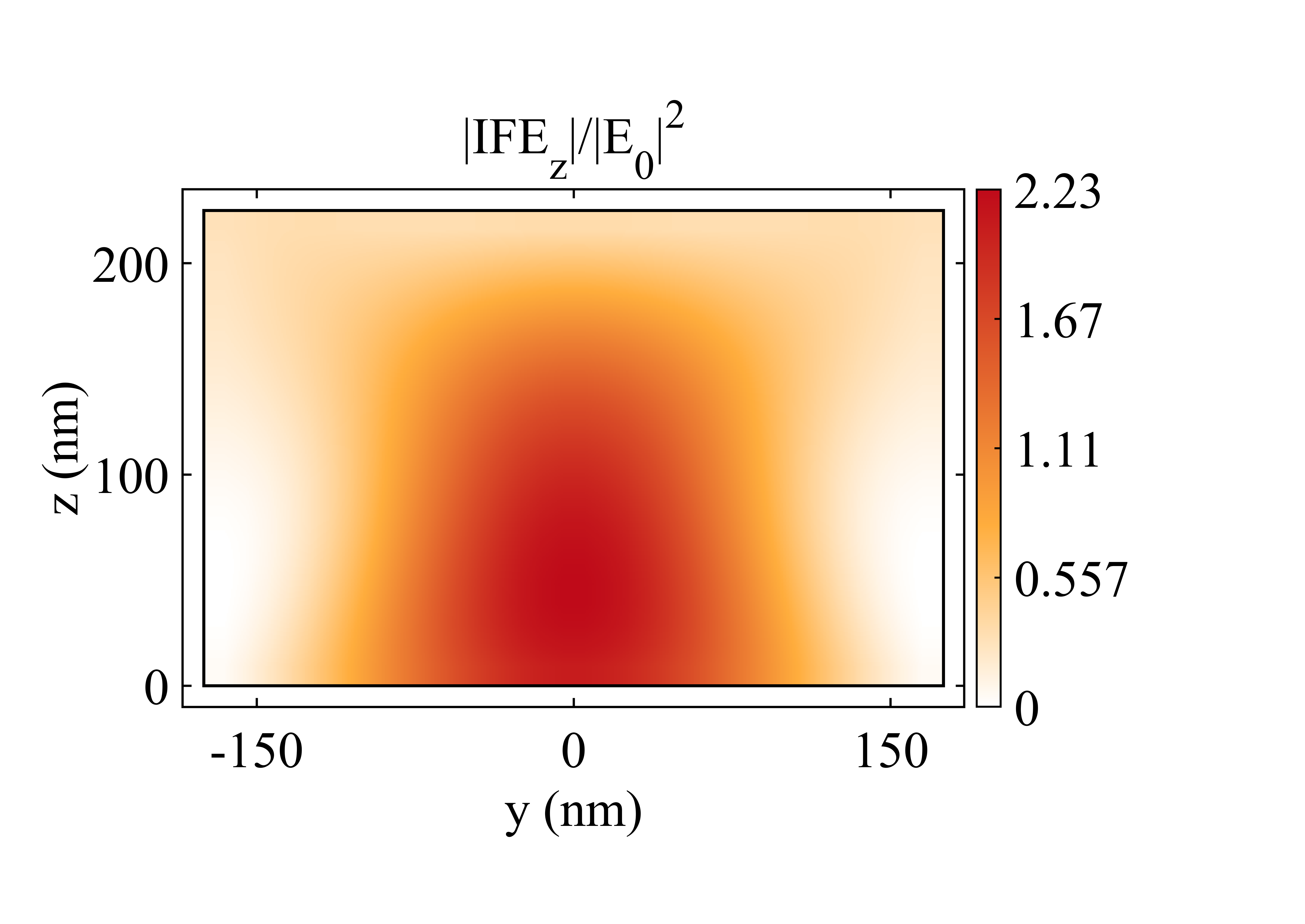}
\par\vspace{1.5mm}
(c)
\end{minipage}%
\hspace{0.025\textwidth}%
\begin{minipage}[b]{0.43\textwidth}
\centering
\includegraphics[width=\linewidth,height=50mm,keepaspectratio,trim={22.40346pt 52.39716pt 129.53783pt 61.06754pt},clip]{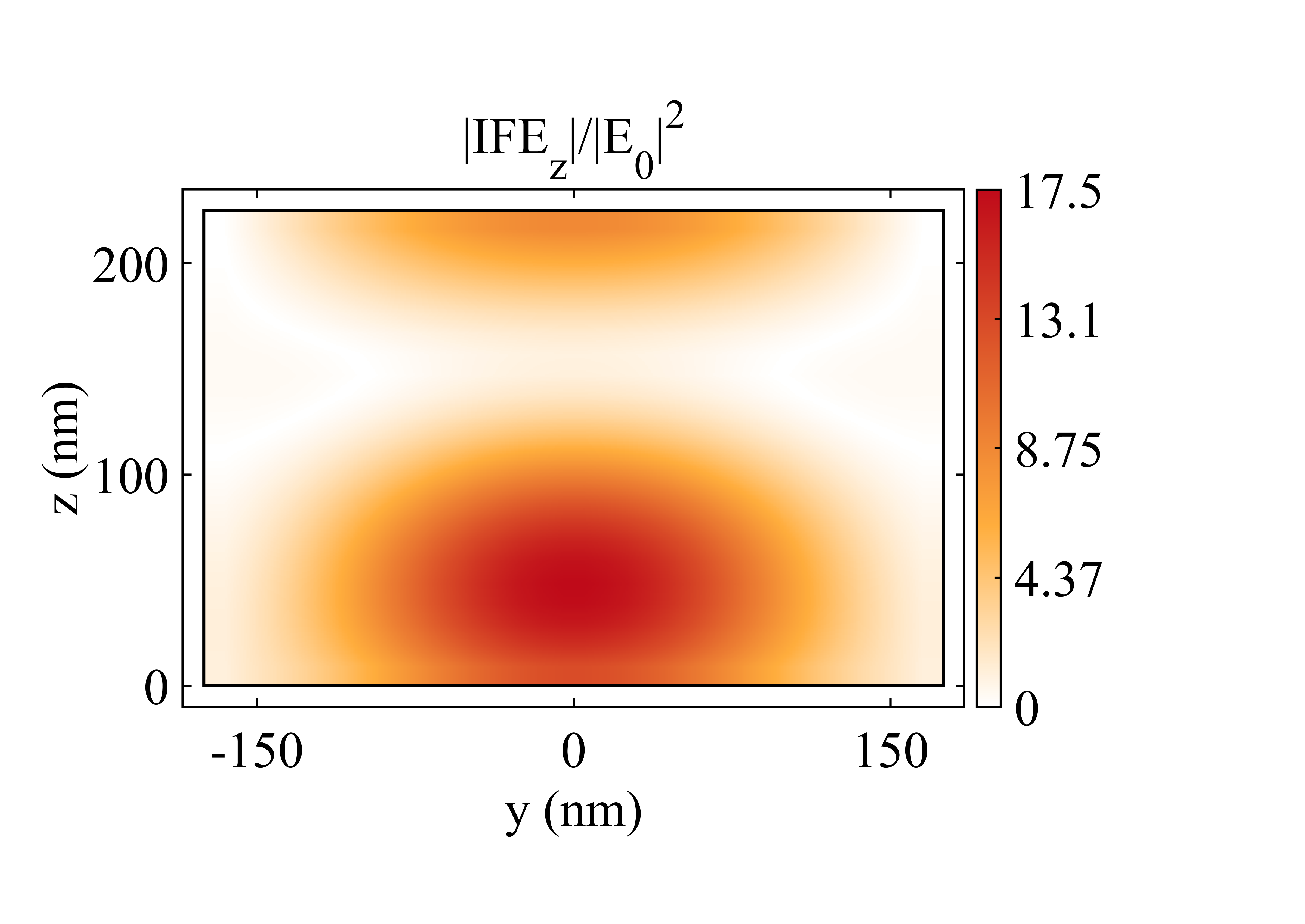}
\par\vspace{1.5mm}
(d)
\end{minipage}
\par\smallskip
\caption{Spatial distributions of the normalized axial IFE-related component, $|\mathrm{IFE}_z|/|E_0|^2$, inside the BIG nanodisk under normally incident circularly polarized radiation. Panels (a,c) correspond to the nanodisk on homogeneous GGG at $\lambda=800$~nm, and panels (b,d) to the nanodisk on the SGG/BIG photonic-crystal substrate at $\lambda=837$~nm, corresponding to the maximum of the summed multipolar scattering contributions. The upper row shows horizontal $xy$-cross-sections through the axial IFE hotspot $z\simeq48.48$~nm in both panels, and the lower row shows vertical $yz$-cross-sections near the disk center. The black contour marks the boundary of the BIG nanodisk. Note the different color scales in the individual panels.}
\label{fig:IFE_maps}
\end{IFEplacedfigure}
\begin{multicols}{2}

Figure~\ref{fig:IFE_maps} shows the spatial distributions of the normalized axial IFE-related component $|\mathrm{IFE}_z|/|E_0|^2$ for the reference nanodisk on homogeneous GGG at $\lambda=800$~nm and the nanodisk on the SGG/BIG photonic crystal at $\lambda=837$~nm. The horizontal $xy$-cross-sections pass through the axial IFE hotspot. The corresponding maps in the MD- and ED-dominated regimes at $811$ and $849$~nm are presented in Appendix~B (Fig.~\ref{fig:IFE_z_appendix}).

For the reference nanodisk on homogeneous GGG, Fig.~\ref{fig:IFE_maps}(a,c), the maximum normalized axial component in the examined sections is approximately $2.2$. For the nanodisk on the photonic crystal at $837$~nm, it reaches approximately $17.5$, Fig.~\ref{fig:IFE_maps}(b,d), corresponding to a $7.9$-fold enhancement. Strong axial enhancement is also obtained in the MD- and ED-dominated regimes, as summarized in Table~\ref{tab:IFE_enhancement} and illustrated in Appendix~B (Fig.~\ref{fig:IFE_z_appendix}).

\end{multicols}
\begin{IFEplacedtable}
\centering
\caption{Largest normalized axial IFE-related values found in the examined disk cross-sections. Enhancements are relative to the disk on GGG at $800$~nm.}
\label{tab:IFE_enhancement}
{\begin{tabular}{lccc}
\hline
Structure and regime & $\lambda$ (nm) & Max. $|\mathrm{IFE}_z|/|E_0|^2$ & Enhancement \\
\hline
GGG, reference & 800 & 2.2 & -- \\
SGG/BIG PC, maximum of multipolar sum & 837 & 17.5 & 7.9 \\
SGG/BIG PC, MD-dominated regime & 811 & 16.3 & 7.3 \\
SGG/BIG PC, ED-dominated regime & 849 & 14 & 6.3 \\
\hline
\end{tabular}
}
\end{IFEplacedtable}
\begin{multicols}{2}

\subsubsection{In-plane IFE components for out-of-plane magnetization}

We assume that an external magnetic field applied along the disk axis aligns the equilibrium magnetization uniformly along $z$. The axial IFE effective field is then parallel to the magnetization and does not contribute to the initial excitation torque. The $x$ and $y$ components are perpendicular to the magnetization and can exert a torque on it. We therefore examine their normalized resultant magnitude, $\sqrt{\mathrm{IFE}_x^2+\mathrm{IFE}_y^2}/|E_0|^2$. Figure~\ref{fig:IFE_inplane} compares this quantity for the nanodisk on GGG at $800$~nm and on the photonic crystal in the MD-dominated regime at $811$~nm.

\end{multicols}
\begin{IFEplacedfigure}
\centering
\begin{minipage}[b]{0.43\textwidth}
\centering
\includegraphics[width=\linewidth,height=57mm,keepaspectratio,trim={10.48053pt 60.52994pt 114.91536pt 63.78085pt},clip]{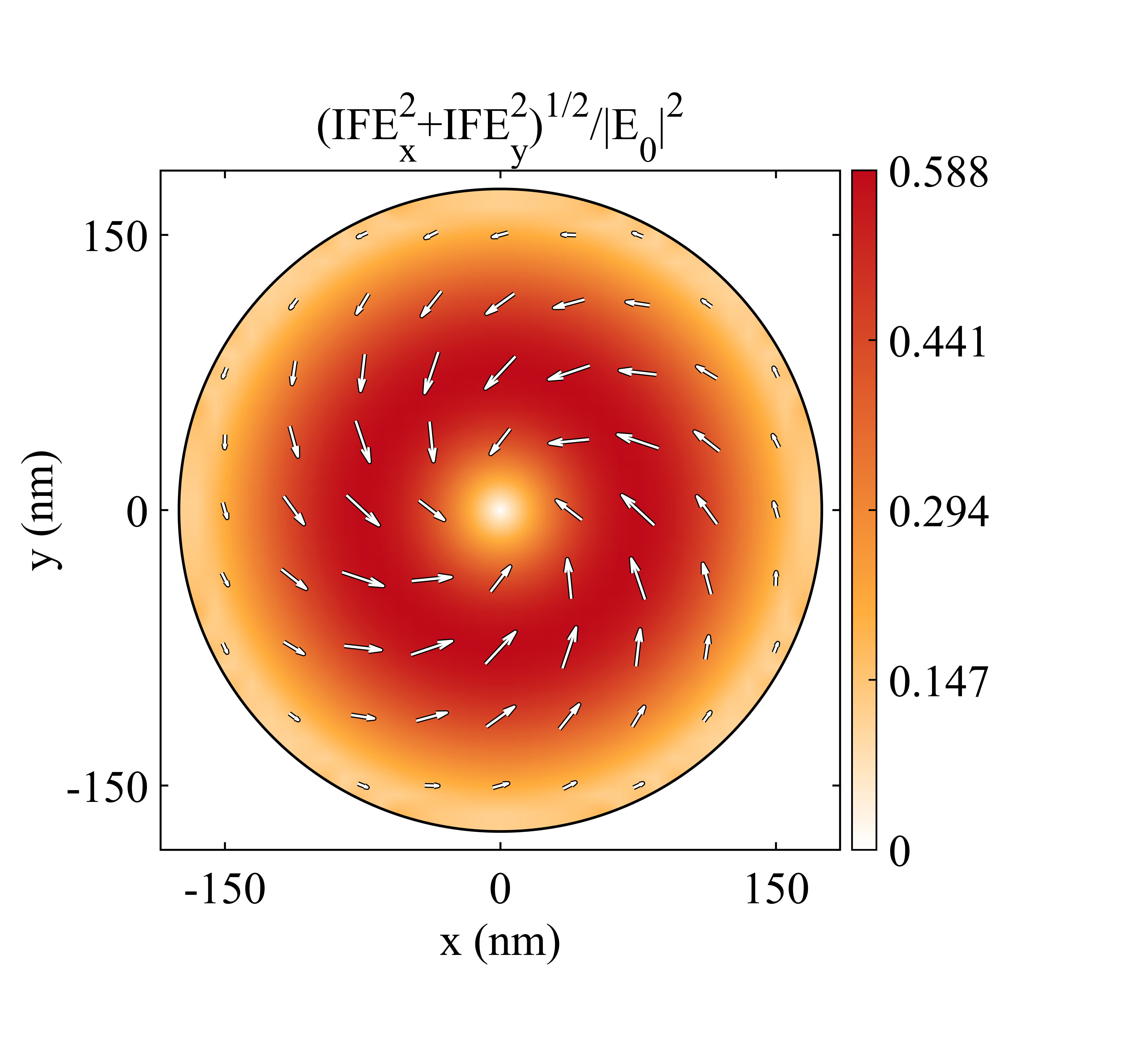}
\par\vspace{1.5mm}
(a)
\end{minipage}%
\hspace{0.025\textwidth}%
\begin{minipage}[b]{0.43\textwidth}
\centering
\includegraphics[width=\linewidth,height=57mm,keepaspectratio,trim={2.96419pt 52.9364pt 124.49606pt 56.33287pt},clip]{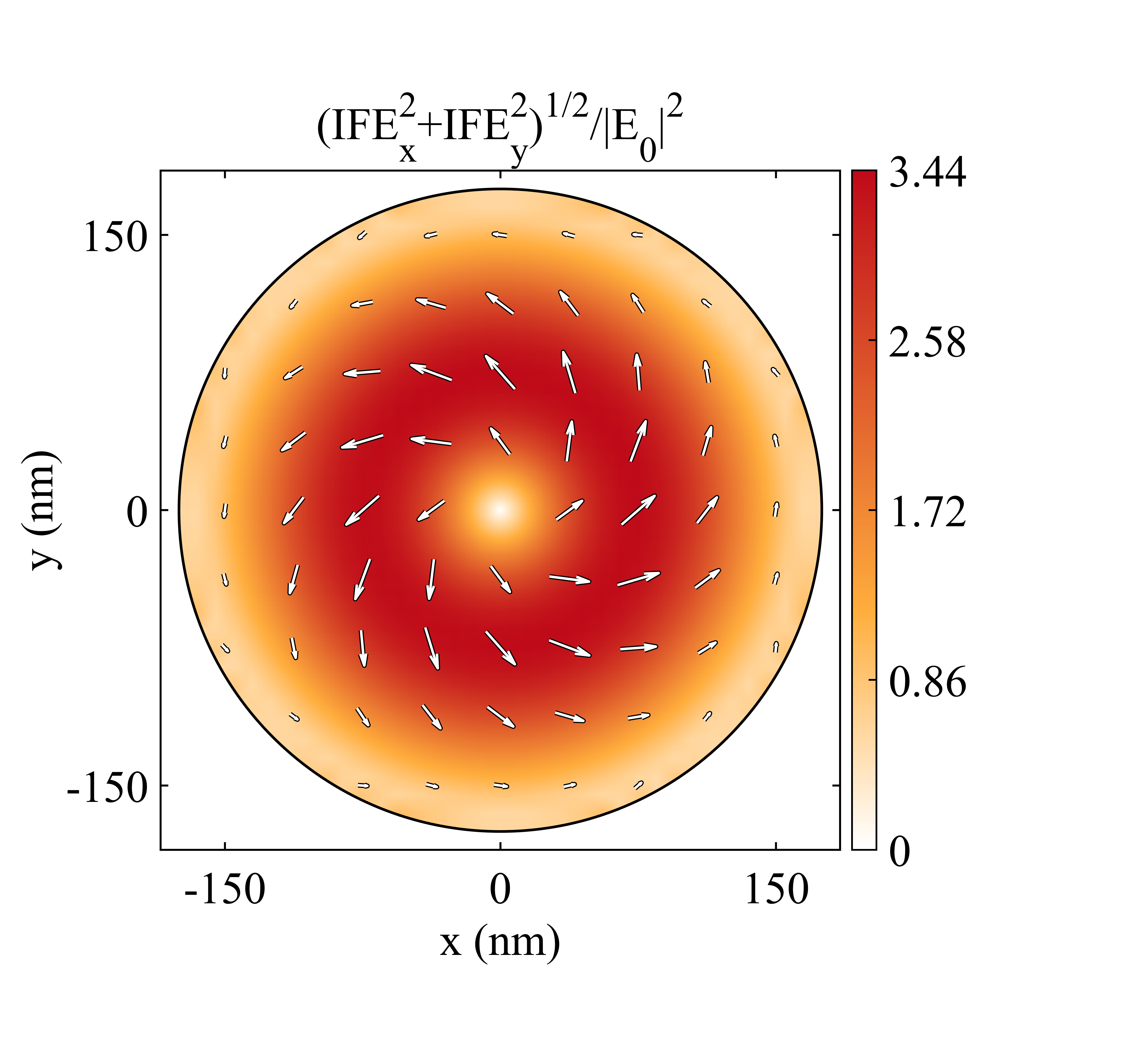}
\par\vspace{1.5mm}
(b)
\end{minipage}
\par\smallskip
\caption{Spatial distributions of the normalized in-plane IFE-related magnitude, $\sqrt{\mathrm{IFE}_x^2+\mathrm{IFE}_y^2}/|E_0|^2$, in horizontal $xy$-cross-sections of the BIG nanodisk under normally incident circularly polarized radiation: (a) the reference nanodisk on homogeneous GGG at $\lambda=800$~nm; (b) the nanodisk on the SGG/BIG photonic-crystal substrate at $\lambda=811$~nm, in the MD-dominated regime. The cross-sections pass through the axial IFE hotspot of the corresponding structure ($z\simeq50$~nm in panel (a) and $z\simeq40$~nm in panel (b)). The white arrows show the direction of $(\mathrm{IFE}_x,\mathrm{IFE}_y)$, the in-plane projection of $\mathrm{Im}[\mathbf{E}\times\mathbf{E}^{*}]$. Arrow lengths are scaled separately in the two panels; amplitudes should be compared using the color bars. The black contour marks the disk boundary. Note the different color scales.}
\label{fig:IFE_inplane}
\end{IFEplacedfigure}
\begin{multicols}{2}

A distinctive feature of the transverse IFE field is its vortex-like vector pattern around a central minimum in magnitude. The arrows in Fig.~\ref{fig:IFE_inplane} reveal this spatial variation of the local field direction. Related vortex-like in-plane IFE fields were previously analyzed for garnet nanospheres under Mie-resonant excitation, with representative distributions shown for the MD and MQ resonances~\cite{krichevsky2024inverse}.

The photonic crystal substantially enhances this spatially structured transverse field: the maximum normalized magnitude in the horizontal maps of Fig.~\ref{fig:IFE_inplane} increases from approximately $0.59$ for the reference nanodisk on GGG to $3.4$ for the nanodisk on the photonic crystal. Vortex-like distributions occur in both the MD- and ED-dominated regimes, with a larger peak magnitude at $811$~nm than at $849$~nm in the examined sections; we therefore use the MD-dominated regime for the main comparison. The corresponding vertical sections and additional maps at $837$ and $849$~nm are discussed in Appendix~B (Figs.~\ref{fig:IFE_inplane_yz} and~\ref{fig:IFE_inplane_appendix}). This spatially nonuniform effective field may couple to nonuniform magnetic modes with a suitable spatial overlap; the subsequent magnetization dynamics is not simulated here.

\section{Conclusion}

We have shown that an all-garnet SGG/BIG photonic crystal substrate can control the resonant response of a single BIG nanodisk on GGG. In contrast to a homogeneous GGG substrate, the photonic crystal suppresses downward leakage of the resonant field and increases field localization inside the magnetic volume.

The substrate modifies the spectral positions and relative strengths of the dipolar contributions and the spatial distribution of the optical field inside the nanodisk, producing MD- and ED-dominated regimes at $\lambda=811$~nm and $849$~nm, respectively. Compared with the reference nanodisk on homogeneous GGG at $\lambda=800$~nm, the maximum normalized axial IFE-related component in the examined disk cross-sections is enhanced by factors of approximately 7.3 and 6.3 in the MD- and ED-dominated regimes, respectively. An even larger enhancement of approximately 7.9 is obtained near $\lambda=837$~nm, corresponding to the maximum of the summed multipolar scattering contributions. At the peaks of the spatial-maximum and volume-average spectra for the disk on the photonic crystal, comparison with the same nanodisk on a homogeneous substrate at the same respective wavelengths gives enhancement factors of approximately $9.8$ and $8.2$ for the IFE-related magnitude.
In addition to the axial enhancement, the photonic-crystal substrate strengthens the vortex-like in-plane IFE-related field relative to the nanodisk on homogeneous GGG. This spatially nonuniform field provides a transverse optical drive for an equilibrium magnetization oriented along the disk axis. These results demonstrate that engineering the photonic environment enables spectral control of the magnitude and spatial distribution of the local optomagnetic response without changing the geometry of the magnetic nanodisk. Combined with epitaxial compatibility, this approach provides a basis for developing garnet nanostructures for optically driven magnetization dynamics.

\section*{ACKNOWLEDGMENTS}

This work was supported by the Russian Science Foundation (Grant No. 25-22-00253). P.V.Z. and V.I.B. acknowledge  support from the Foundation for the Advancement of Theoretical Physics and Mathematics ``Basis'' (Grant No. 25-1-1-49-6).
The authors thank A. Kalish for fruitful discussion.

\section*{CONFLICT OF INTEREST}

The authors declare no conflict of interest.

\section*{Appendix}
\subsection*{Appendix A. Material Parameters}

The wavelength-dependent dielectric permittivities used in the numerical simulations are shown in Fig.~\ref{fig:epsilon_dispersion}. The material dispersion of BIG, SGG, and GGG was taken from~\cite{ignatyeva2024asymmetric}. In the considered spectral range, the permittivities were calculated as
\begin{equation}
\varepsilon_{\mathrm{SGG}} =
1 + \frac{2.75}{1-\left(128/\lambda\right)^2},
\label{eq:eps_SGG}
\end{equation}

\begin{equation}
\varepsilon_{\mathrm{BIG}} =
1 + \frac{4.70}{1-\left(303/\lambda\right)^2}
  + \frac{0.08}{1-\left(494/\lambda\right)^2}
  + 0.07\left(\frac{494}{\lambda}\right)i,
\label{eq:eps_BIG}
\end{equation}

\begin{equation}
n_{\mathrm{GGG}} =
1.91+\left(\frac{203}{\lambda}\right)^2,
\label{eq:n_GGG}
\end{equation}

\begin{equation}
\varepsilon_{\mathrm{GGG}} = n_{\mathrm{GGG}}^2,
\label{eq:eps_GGG}
\end{equation}
where $\lambda$ is expressed in nanometers.
These dispersion relations were used both for the homogeneous GGG substrate and for the SGG/BIG photonic-crystal multilayer. The imaginary part is included only for BIG, while SGG and GGG were treated as lossless dielectric materials in this wavelength range.

\begin{IFEcolumnfigure}
    \centering
    \includegraphics[width=0.95\linewidth]{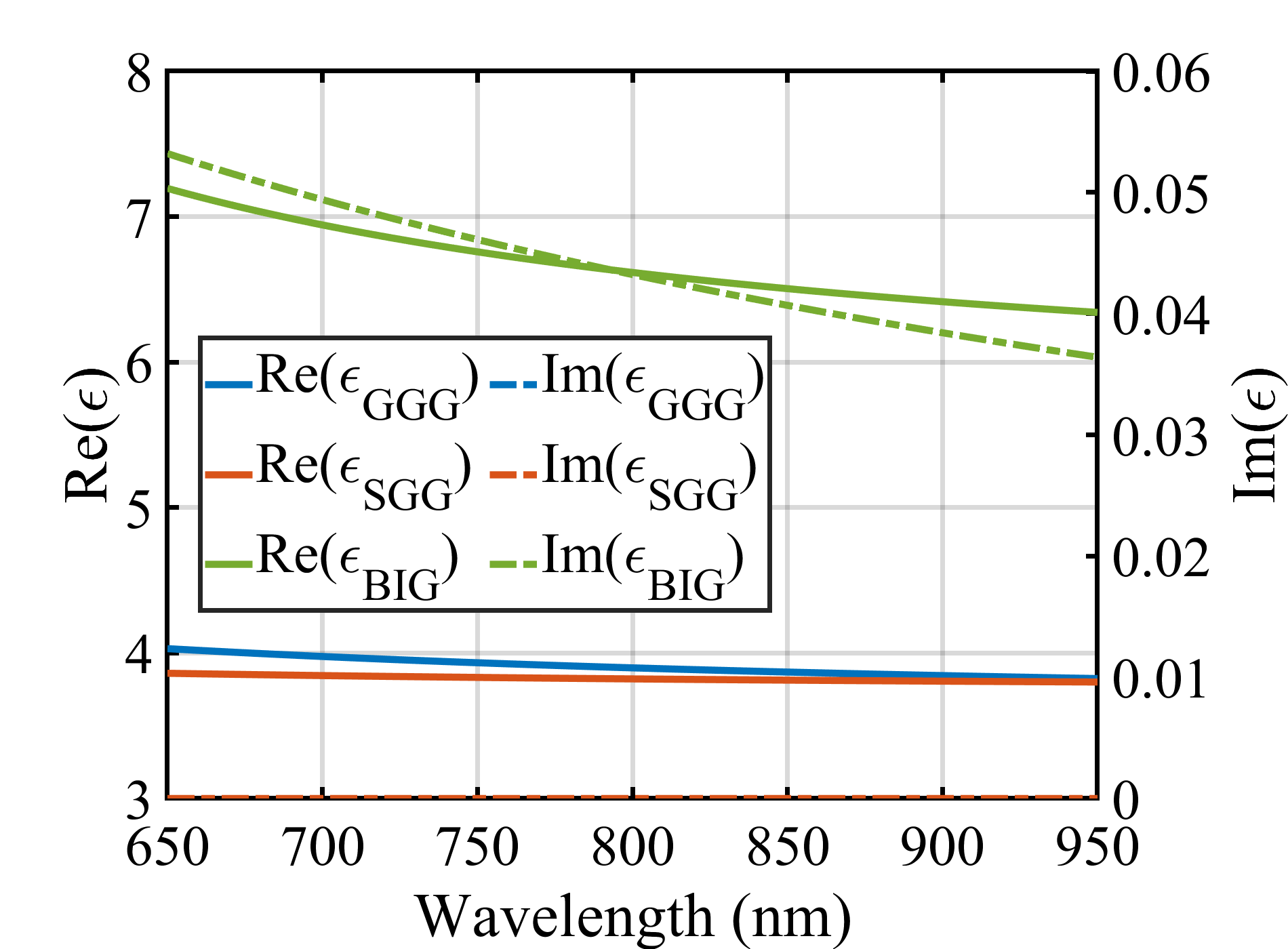}
    \caption{Real and imaginary parts of the dielectric permittivities of BIG, SGG, and GGG used in the numerical simulations. The dispersion relations are taken from~\cite{ignatyeva2024asymmetric}.}
    \label{fig:epsilon_dispersion}
\end{IFEcolumnfigure}


\makeatletter
\makeatother
\subsection*{Appendix B. Additional near-field distributions}

This appendix presents the additional electric-field and IFE-related maps for the dipolar regimes, complementing the comparisons in the main text.

Figure~\ref{fig:E2_maps_appendix} shows the electric-field distributions at $811$ and $849$~nm. The normalized in-disk maxima in the plotted sections are approximately $16$ and $14$, corresponding to enhancements of $7.0$ and $6.2$ relative to the disk on GGG at $800$~nm. Both dipolar regimes therefore provide strong localization inside the magnetic material. The vertical cross-sections show the field distribution in the disk and its penetration into the multilayer.

\end{multicols}
\begin{IFEplacedfigure}
\centering
\begin{minipage}[b]{0.43\textwidth}
\centering
\includegraphics[width=\linewidth,height=37mm,keepaspectratio,trim={10.5405pt 30.85896pt 36.1412pt 27.1039pt},clip]{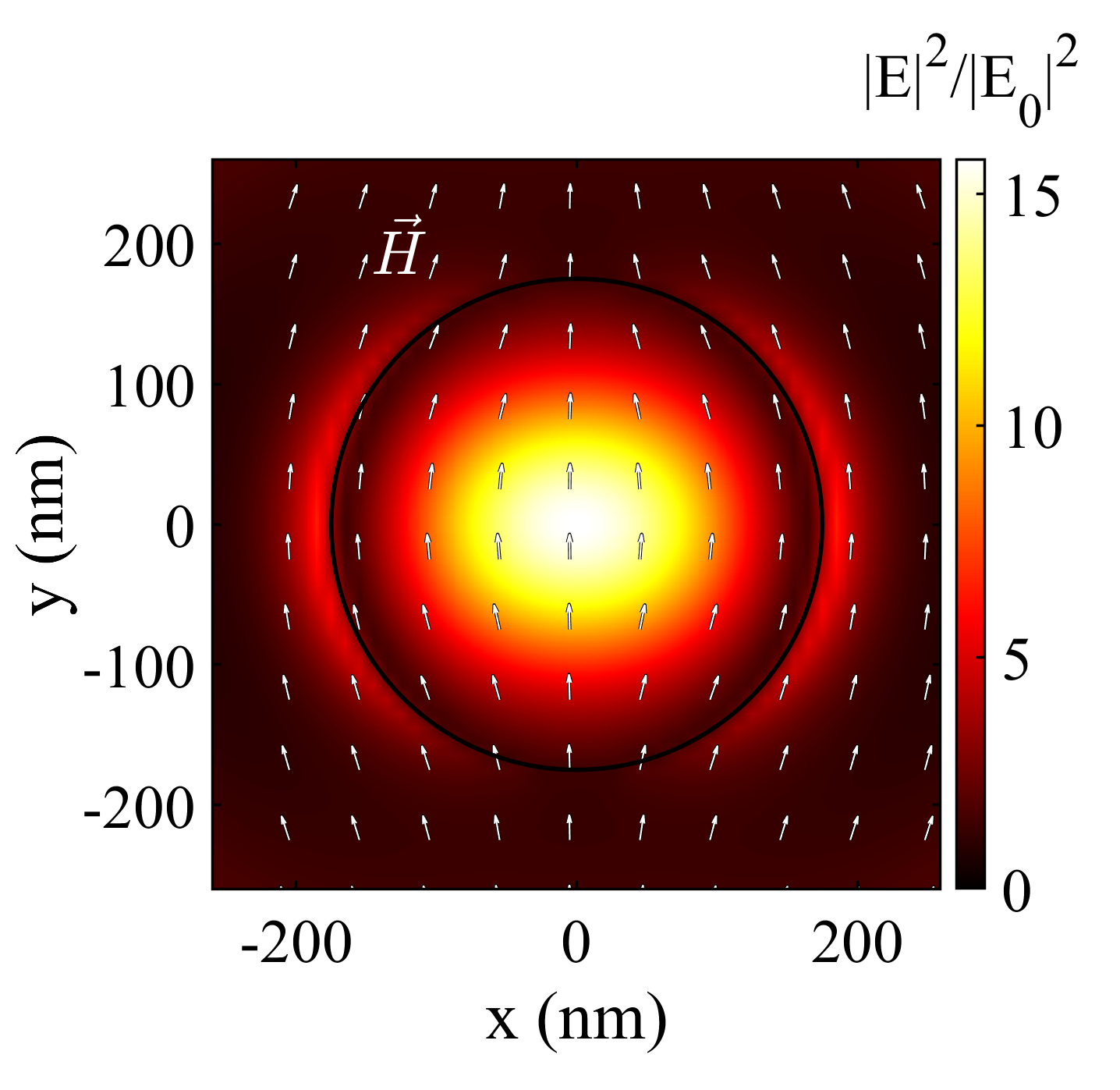}
\par\vspace{1.5mm}
(a)
\end{minipage}%
\hspace{0.025\textwidth}%
\begin{minipage}[b]{0.43\textwidth}
\centering
\includegraphics[width=\linewidth,height=37mm,keepaspectratio,trim={10.5405pt 30.85896pt 36.1412pt 27.1039pt},clip]{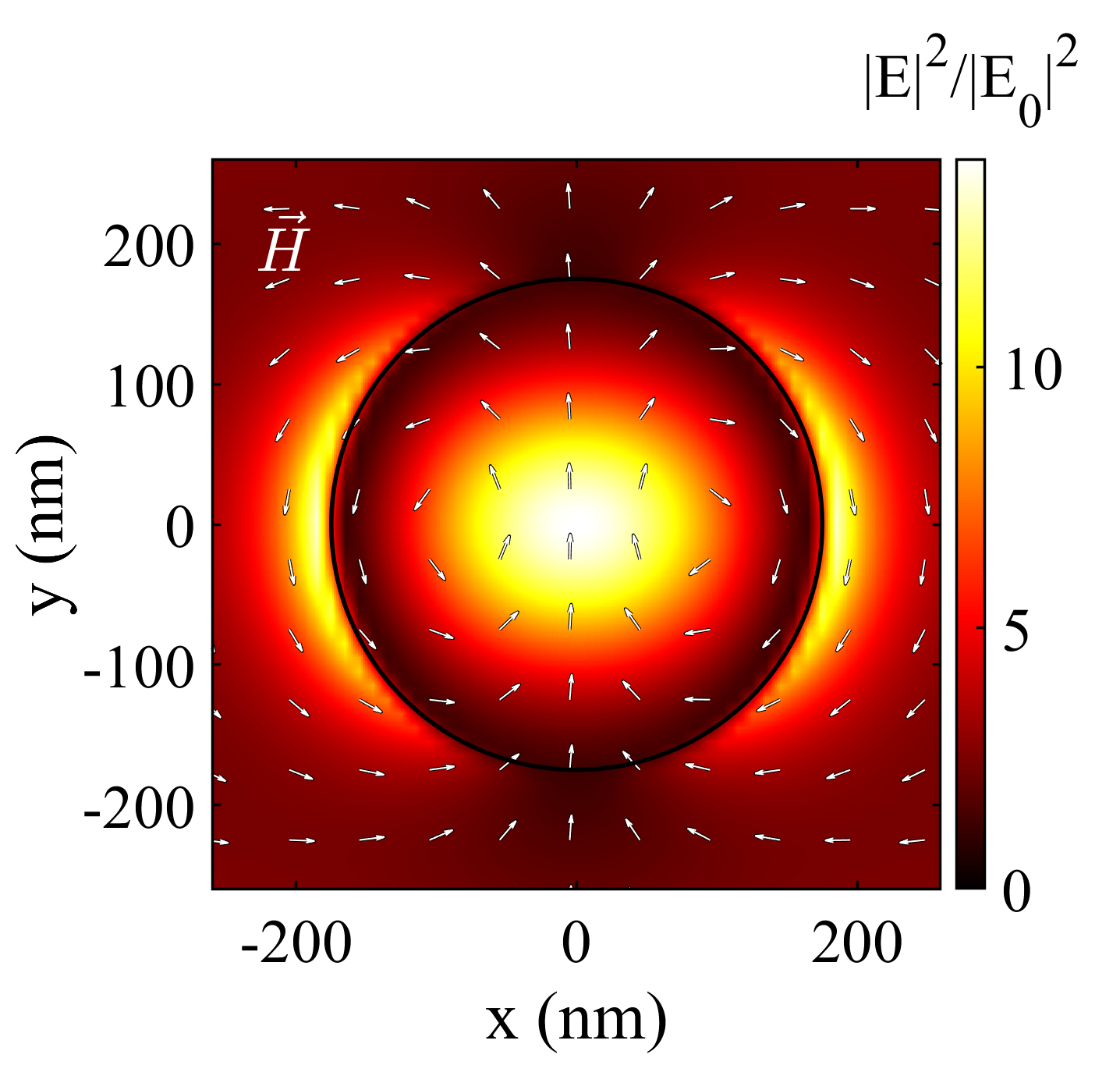}
\par\vspace{1.5mm}
(b)
\end{minipage}
\par\vspace{3mm}
\begin{minipage}[b]{0.43\textwidth}
\centering
\includegraphics[width=\linewidth,height=37mm,keepaspectratio,trim={10.5405pt 0.0pt 30.8627pt 0.0pt},clip]{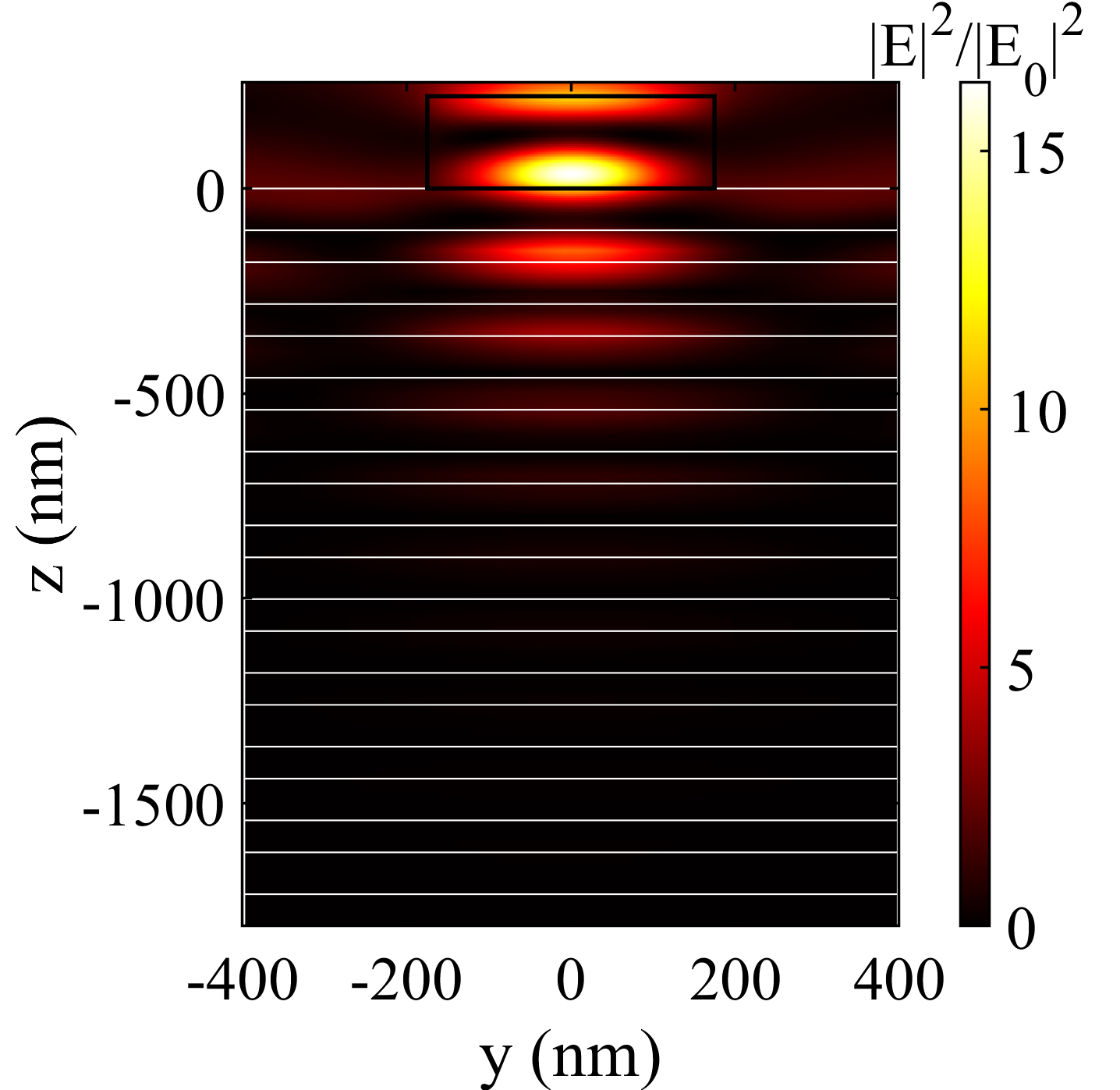}
\par\vspace{1.5mm}
(c)
\end{minipage}%
\hspace{0.025\textwidth}%
\begin{minipage}[b]{0.43\textwidth}
\centering
\includegraphics[width=\linewidth,height=37mm,keepaspectratio,trim={10.5405pt 0.0pt 26.34299pt 0.0pt},clip]{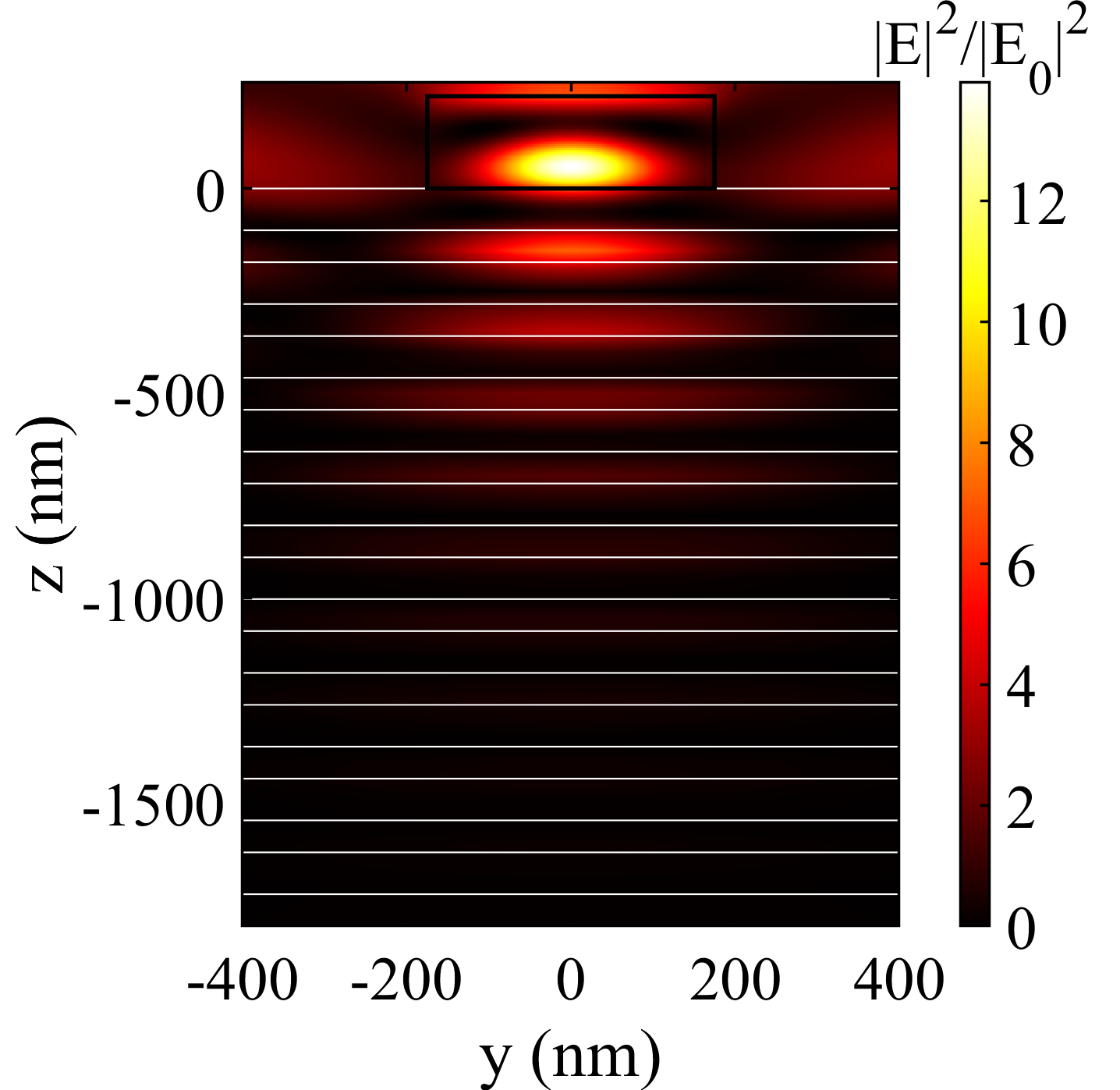}
\par\vspace{1.5mm}
(d)
\end{minipage}
\par\smallskip
\caption{Near-field distributions of the normalized electric-field intensity, $|E|^2/|E_0|^2$, for the BIG nanodisk on the SGG/BIG photonic-crystal substrate under linearly polarized radiation. Panels (a,c) and (b,d) correspond to $\lambda=811$~nm and $849$~nm, respectively, at the maxima of the MD and ED contributions. Panels (a,b) show horizontal $xy$-cross-sections taken at the height of maximum $|E|^2$ inside the nanodisk ($z\simeq50$~nm in both panels), and panels (c,d) show vertical $yz$-cross-sections near the disk center. The black contour marks the disk boundary, and the white horizontal lines indicate the interfaces of the photonic-crystal layers. In the horizontal cross-sections, the white arrows show the instantaneous in-plane optical magnetic field, $(H_x,H_y)$, at a fixed optical phase. The color scales show $|E|^2/|E_0|^2$, with $E_0$ the incident electric-field amplitude. Note the different color scales.}
\label{fig:E2_maps_appendix}
\end{IFEplacedfigure}
\begin{multicols}{2}

Figure~\ref{fig:IFE_z_appendix} shows the axial IFE-related component for the same two dipolar regimes. Its maximum normalized magnitude in the examined sections reaches approximately $16.3$ at $811$~nm and $14$ at $849$~nm. The corresponding enhancement factors, $7.3$ and $6.3$, are given in Table~\ref{tab:IFE_enhancement}. Together with Fig.~\ref{fig:IFE_maps}, these maps show that axial enhancement is maintained across the three examined wavelengths, with the largest value at $837$~nm.

\end{multicols}
\begin{IFEplacedfigure}
\centering
\begin{minipage}[b]{0.43\textwidth}
\centering
\includegraphics[width=\linewidth,height=37mm,keepaspectratio,trim={10.48053pt 60.52994pt 132.2506pt 64.31458pt},clip]{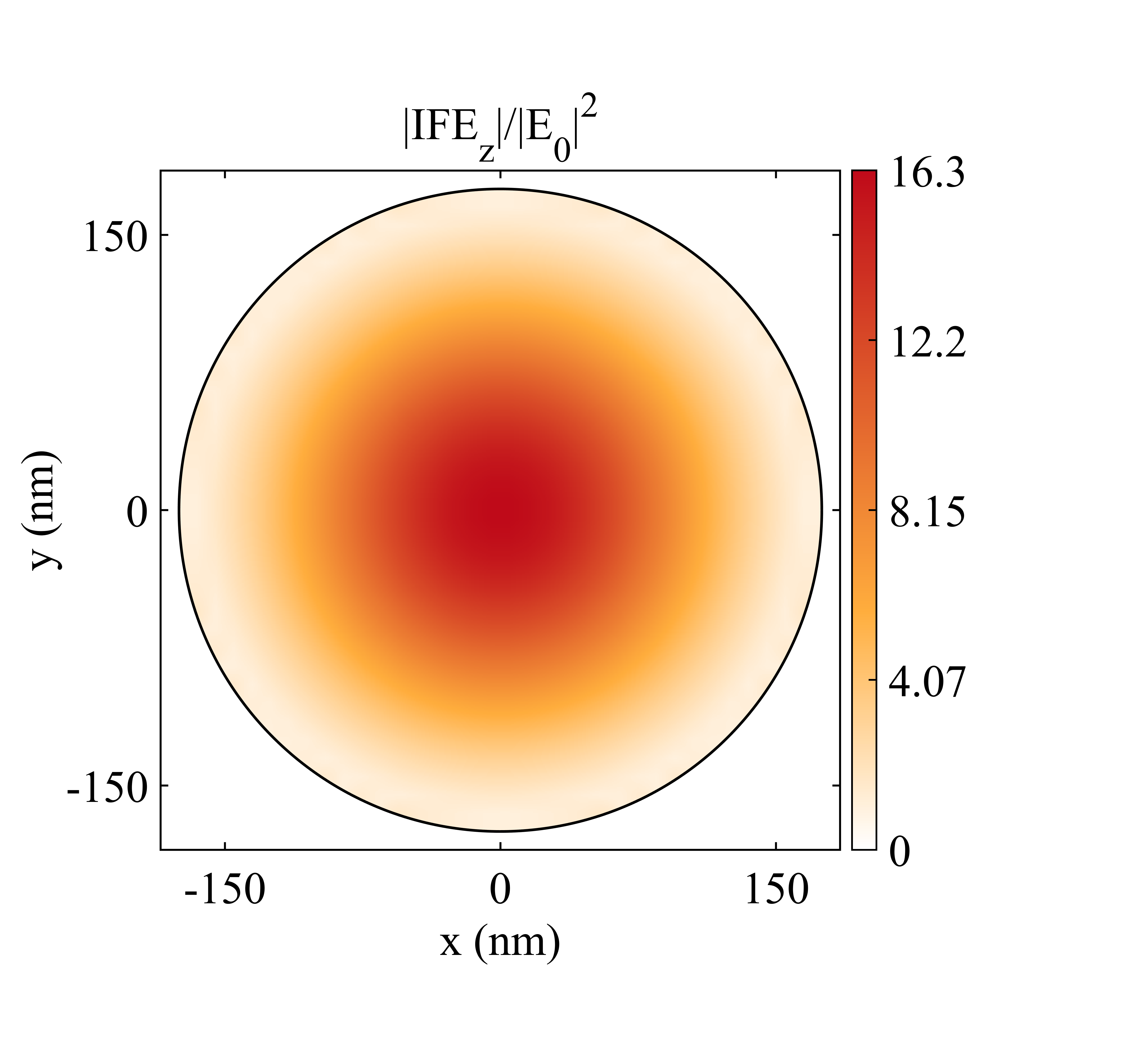}
\par\vspace{1.5mm}
(a)
\end{minipage}%
\hspace{0.025\textwidth}%
\begin{minipage}[b]{0.43\textwidth}
\centering
\includegraphics[width=\linewidth,height=37mm,keepaspectratio,trim={10.48053pt 60.52994pt 132.2506pt 64.31458pt},clip]{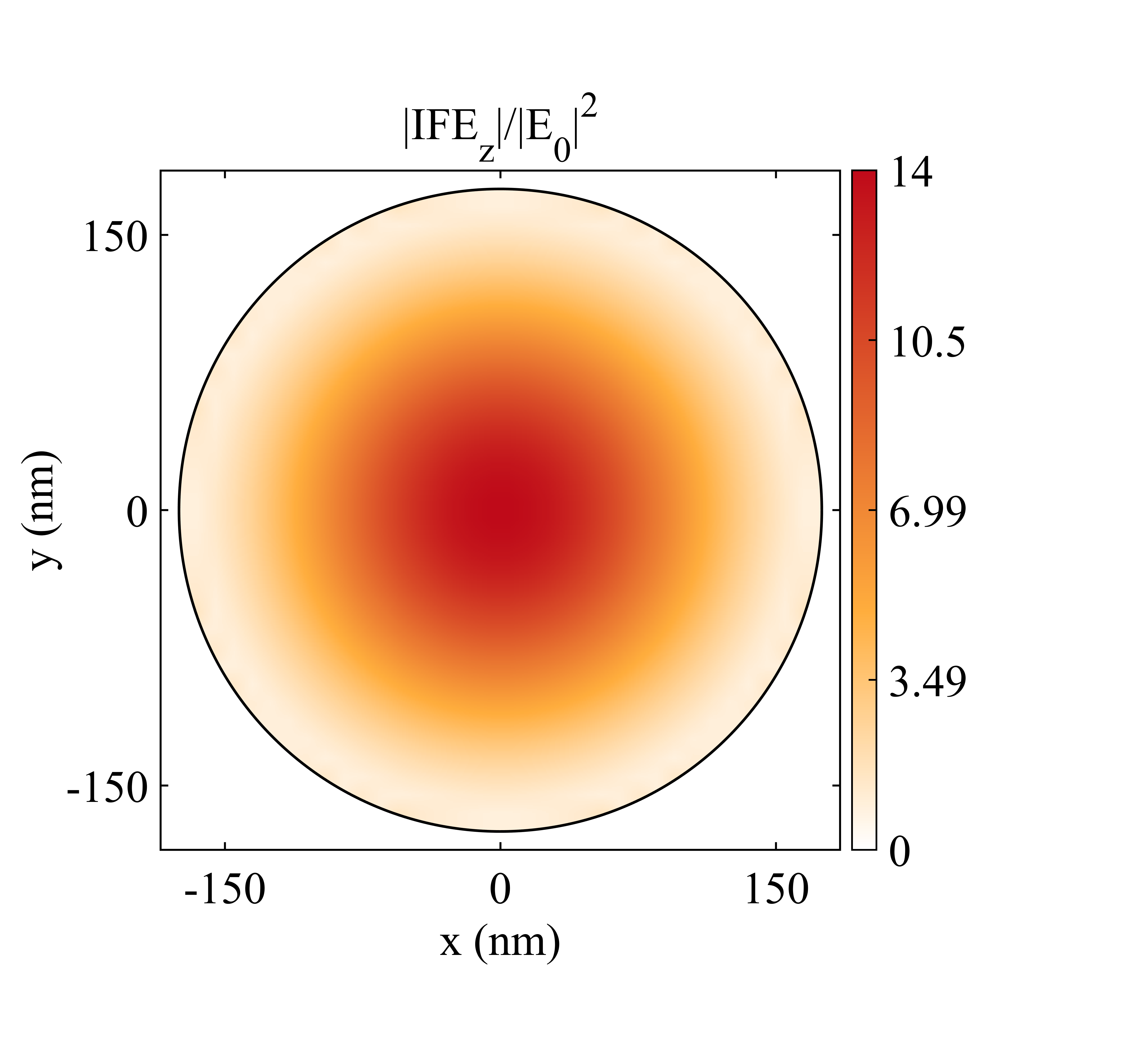}
\par\vspace{1.5mm}
(b)
\end{minipage}
\par\vspace{3mm}
\begin{minipage}[b]{0.43\textwidth}
\centering
\includegraphics[width=\linewidth,height=37mm,keepaspectratio,trim={22.40346pt 52.39716pt 129.53783pt 61.06754pt},clip]{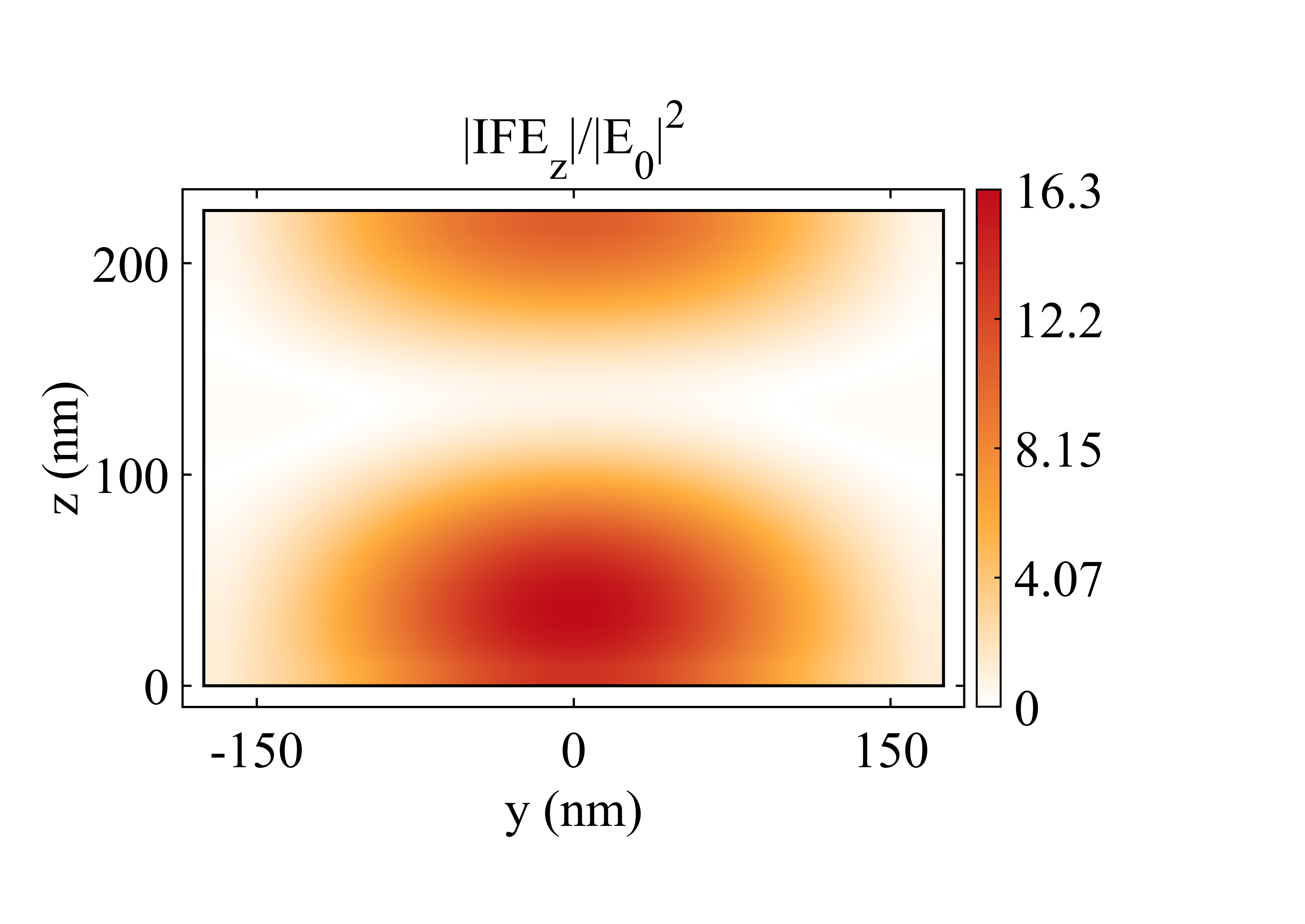}
\par\vspace{1.5mm}
(c)
\end{minipage}%
\hspace{0.025\textwidth}%
\begin{minipage}[b]{0.43\textwidth}
\centering
\includegraphics[width=\linewidth,height=37mm,keepaspectratio,trim={22.40346pt 52.39716pt 129.53783pt 61.06754pt},clip]{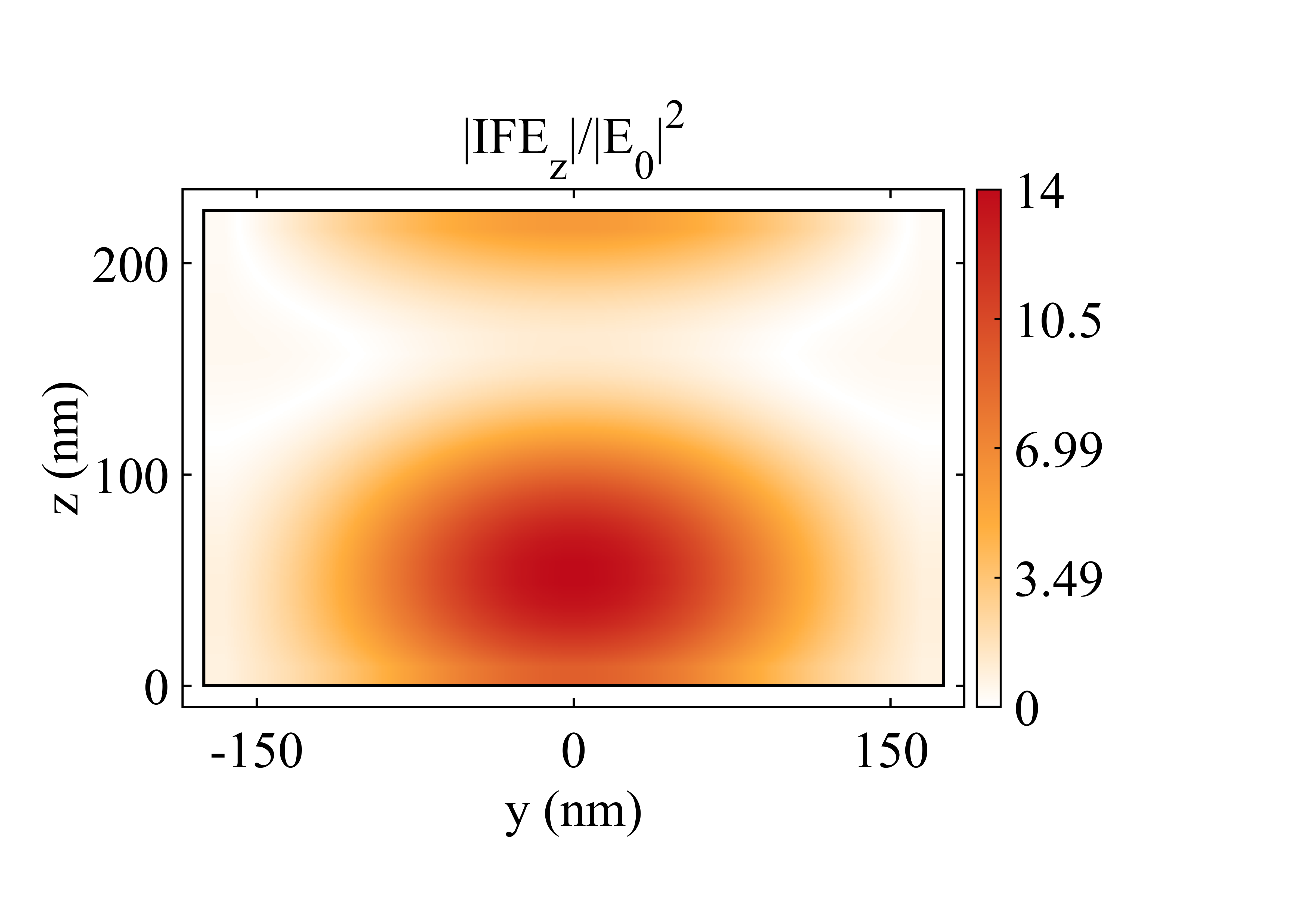}
\par\vspace{1.5mm}
(d)
\end{minipage}
\par\smallskip
\caption{Normalized axial IFE-related component, $|\mathrm{IFE}_z|/|E_0|^2$, inside the BIG nanodisk on the SGG/BIG photonic-crystal substrate under circularly polarized excitation. Panels (a,c) correspond to the MD-dominated regime at $\lambda=811$~nm, and panels (b,d) to the ED-dominated regime at $\lambda=849$~nm. Panels (a,b) show horizontal $xy$-cross-sections through the axial IFE hotspot ($z\simeq40$~nm and $50$~nm, respectively), and panels (c,d) show vertical $yz$-cross-sections near the disk center. The black contour marks the disk boundary. Note the different color scales.}
\label{fig:IFE_z_appendix}
\end{IFEplacedfigure}
\begin{multicols}{2}

Figure~\ref{fig:IFE_inplane_yz} complements the horizontal maps in Fig.~\ref{fig:IFE_inplane} by showing how the in-plane IFE-related magnitude varies through the disk height. The plotted quantity remains $\sqrt{\mathrm{IFE}_x^2+\mathrm{IFE}_y^2}/|E_0|^2$; only the sampling plane changes to $yz$. Its maximum in these vertical sections increases from approximately $0.63$ for the disk on GGG at $800$~nm to $4.2$ for the disk on the photonic crystal at $811$~nm. The regions of enhanced transverse drive are spatially nonuniform within the disk.

\end{multicols}
\begin{IFEplacedfigure}
\centering
\begin{minipage}[b]{0.43\textwidth}
\centering
\includegraphics[width=\linewidth,height=57mm,keepaspectratio,trim={22.40346pt 52.39716pt 112.91718pt 60.52388pt},clip]{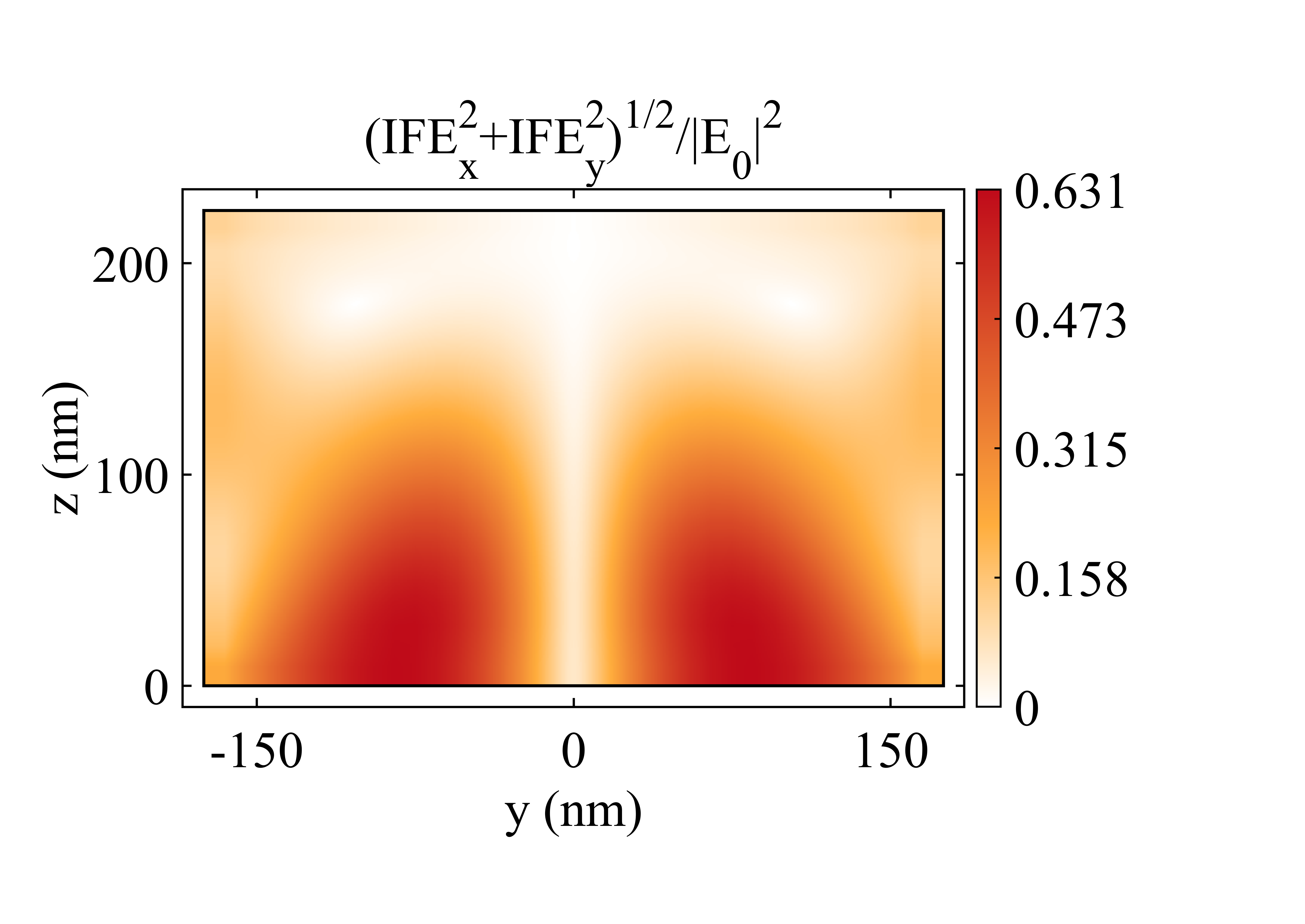}
\par\vspace{1.5mm}
(a)
\end{minipage}%
\hspace{0.025\textwidth}%
\begin{minipage}[b]{0.43\textwidth}
\centering
\includegraphics[width=\linewidth,height=57mm,keepaspectratio,trim={22.40346pt 52.39716pt 129.3658pt 60.52388pt},clip]{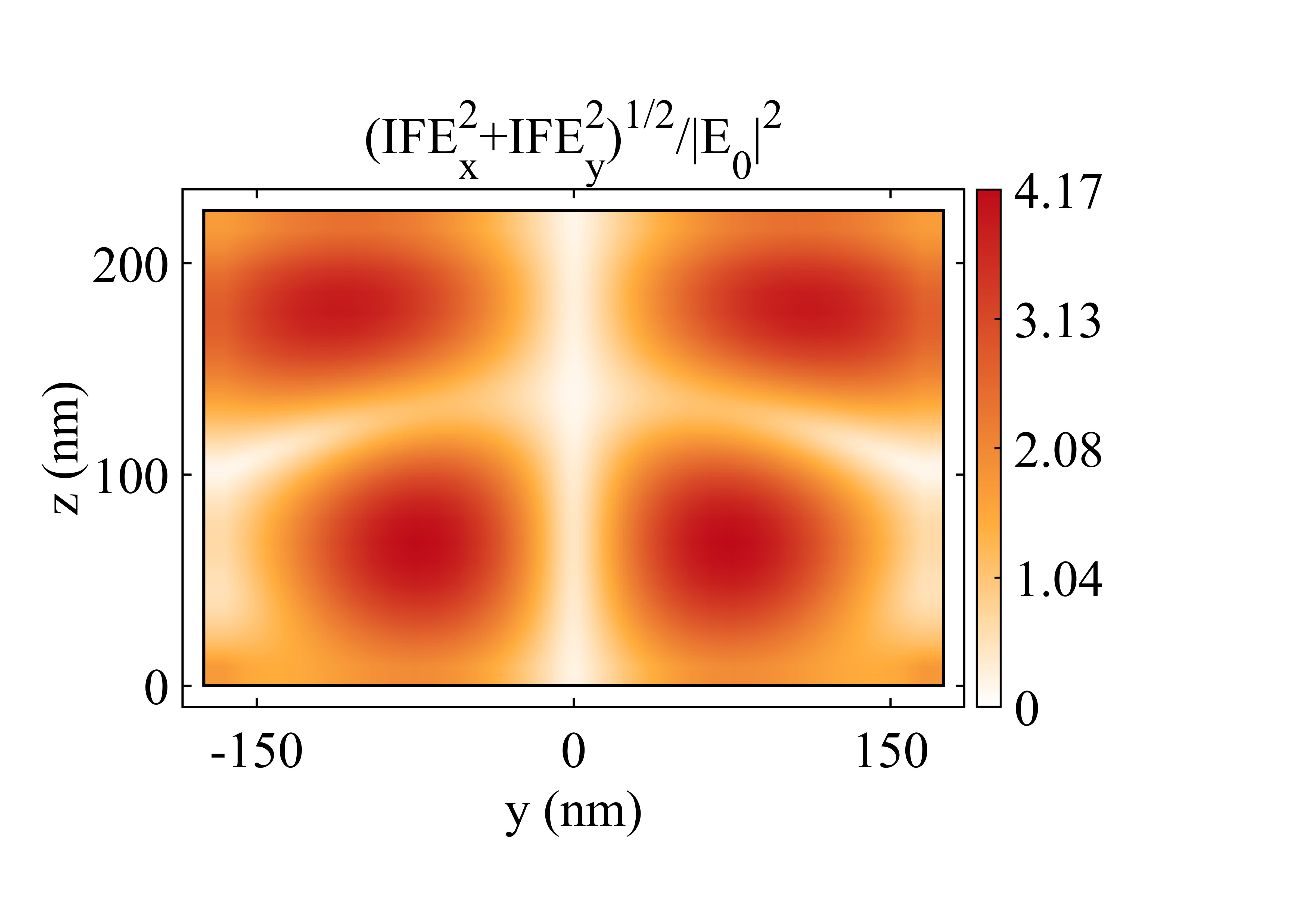}
\par\vspace{1.5mm}
(b)
\end{minipage}
\par\smallskip
\caption{Normalized in-plane IFE-related magnitude, $\sqrt{\mathrm{IFE}_x^2+\mathrm{IFE}_y^2}/|E_0|^2$, shown in vertical $yz$-cross-sections near the disk center: (a) the nanodisk on homogeneous GGG at $\lambda=800$~nm; (b) the nanodisk on the SGG/BIG photonic-crystal substrate at $\lambda=811$~nm. The black contour marks the disk boundary. Note the different color scales. The same in-plane magnitude as in Fig.~\ref{fig:IFE_inplane} is sampled in the $yz$ plane.}
\label{fig:IFE_inplane_yz}
\end{IFEplacedfigure}
\begin{multicols}{2}

Figure~\ref{fig:IFE_inplane_appendix} shows the in-plane IFE-related response at $837$ and $849$~nm. The horizontal cross-sections retain a central minimum and vortex-like vector patterns, demonstrating that this response is also present at the maximum of the multipolar sum and in the ED-dominated regime. Its peak magnitude in the examined sections is lower than at $811$~nm. The vertical cross-sections show the accompanying variation of the in-plane magnitude along the disk height.

\end{multicols}
\begin{IFEplacedfigure}
\centering
\begin{minipage}[b]{0.43\textwidth}
\centering
\includegraphics[width=\linewidth,height=37mm,keepaspectratio,trim={10.48053pt 60.52994pt 115.81522pt 63.78085pt},clip]{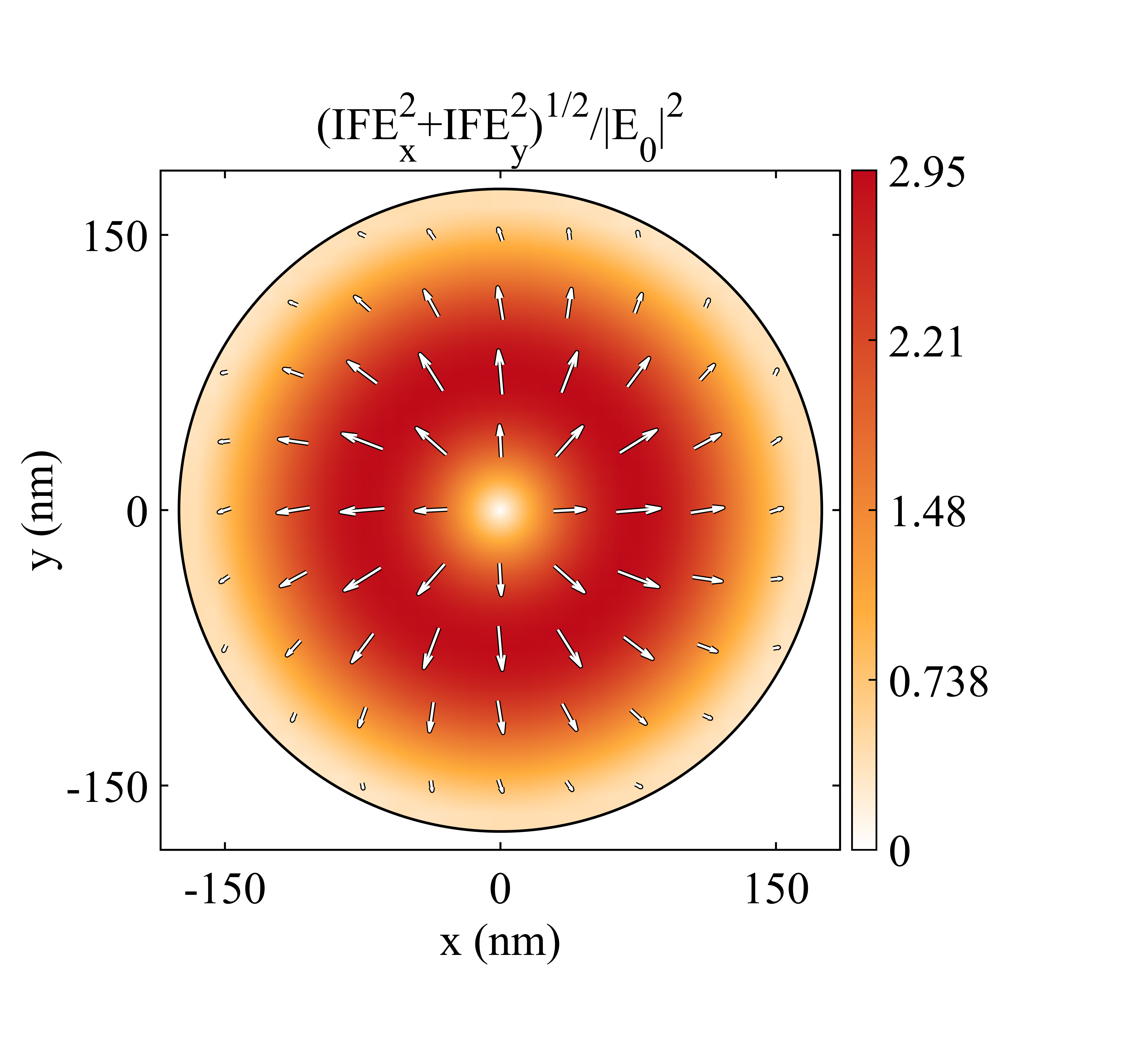}
\par\vspace{1.5mm}
(a)
\end{minipage}%
\hspace{0.025\textwidth}%
\begin{minipage}[b]{0.43\textwidth}
\centering
\includegraphics[width=\linewidth,height=37mm,keepaspectratio,trim={10.48053pt 60.52994pt 115.0874pt 63.78085pt},clip]{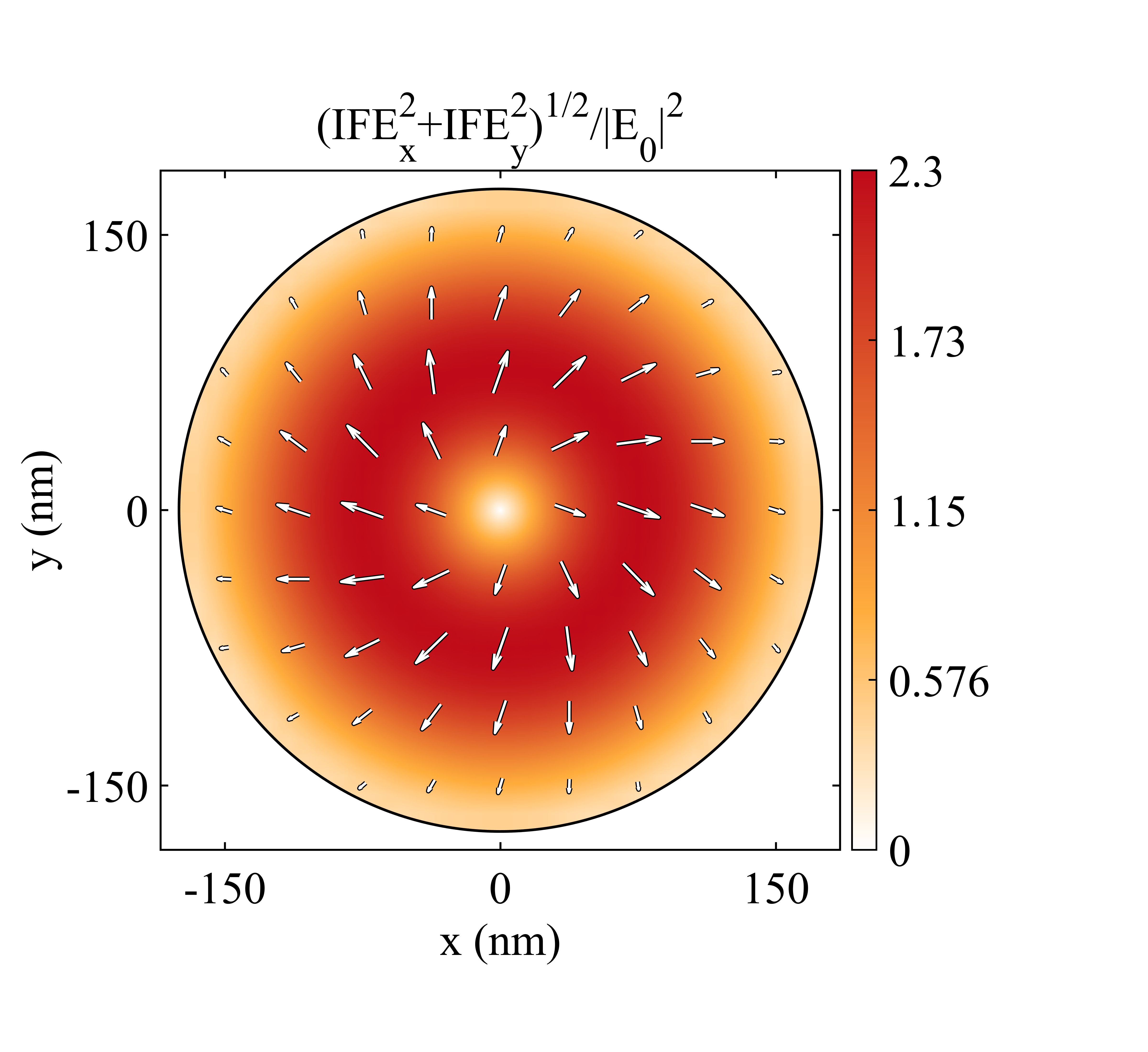}
\par\vspace{1.5mm}
(b)
\end{minipage}
\par\vspace{3mm}
\begin{minipage}[b]{0.43\textwidth}
\centering
\includegraphics[width=\linewidth,height=37mm,keepaspectratio,trim={22.40346pt 52.39716pt 112.91718pt 60.52388pt},clip]{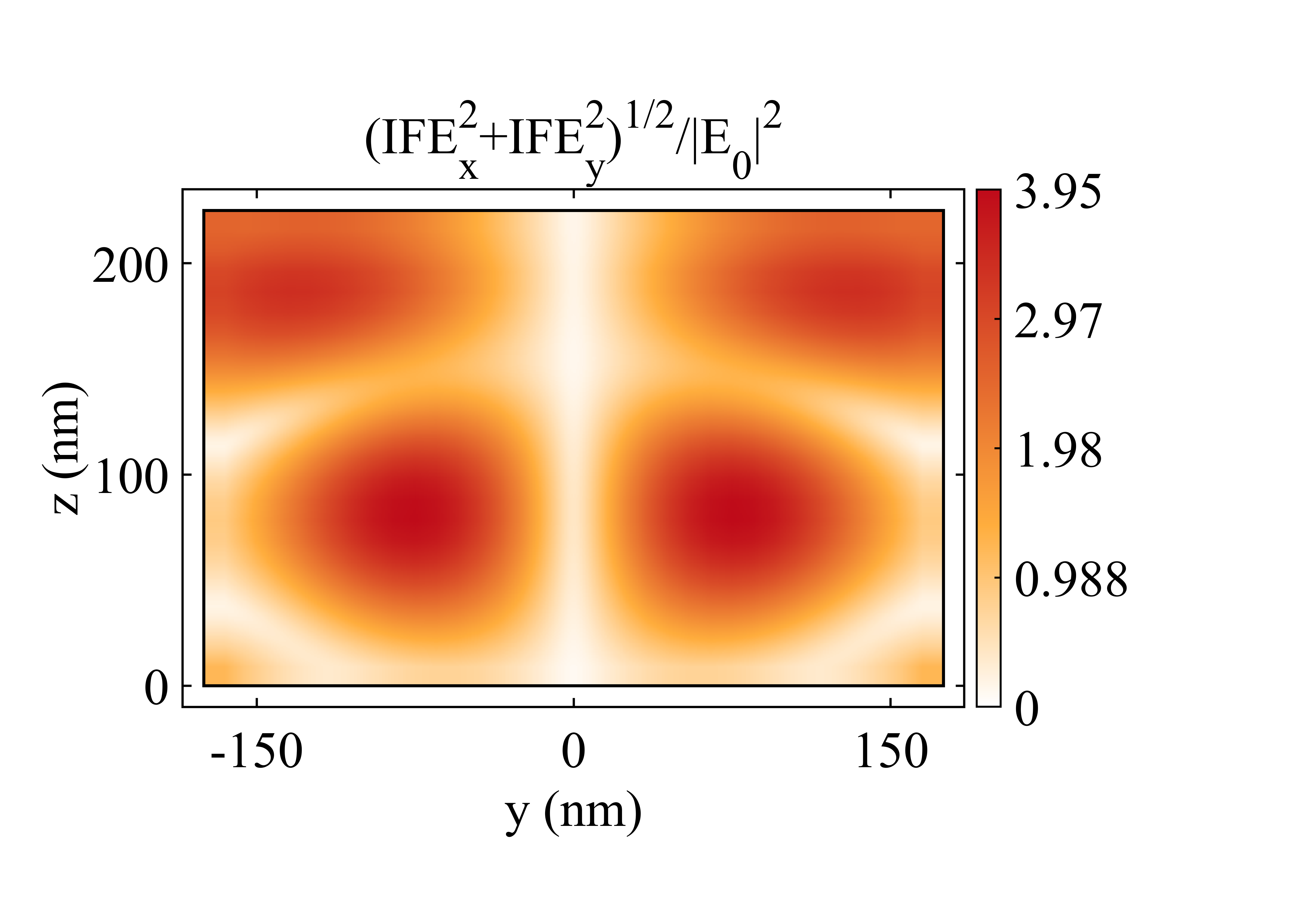}
\par\vspace{1.5mm}
(c)
\end{minipage}%
\hspace{0.025\textwidth}%
\begin{minipage}[b]{0.43\textwidth}
\centering
\includegraphics[width=\linewidth,height=37mm,keepaspectratio,trim={22.40346pt 52.39716pt 113.2877pt 60.52388pt},clip]{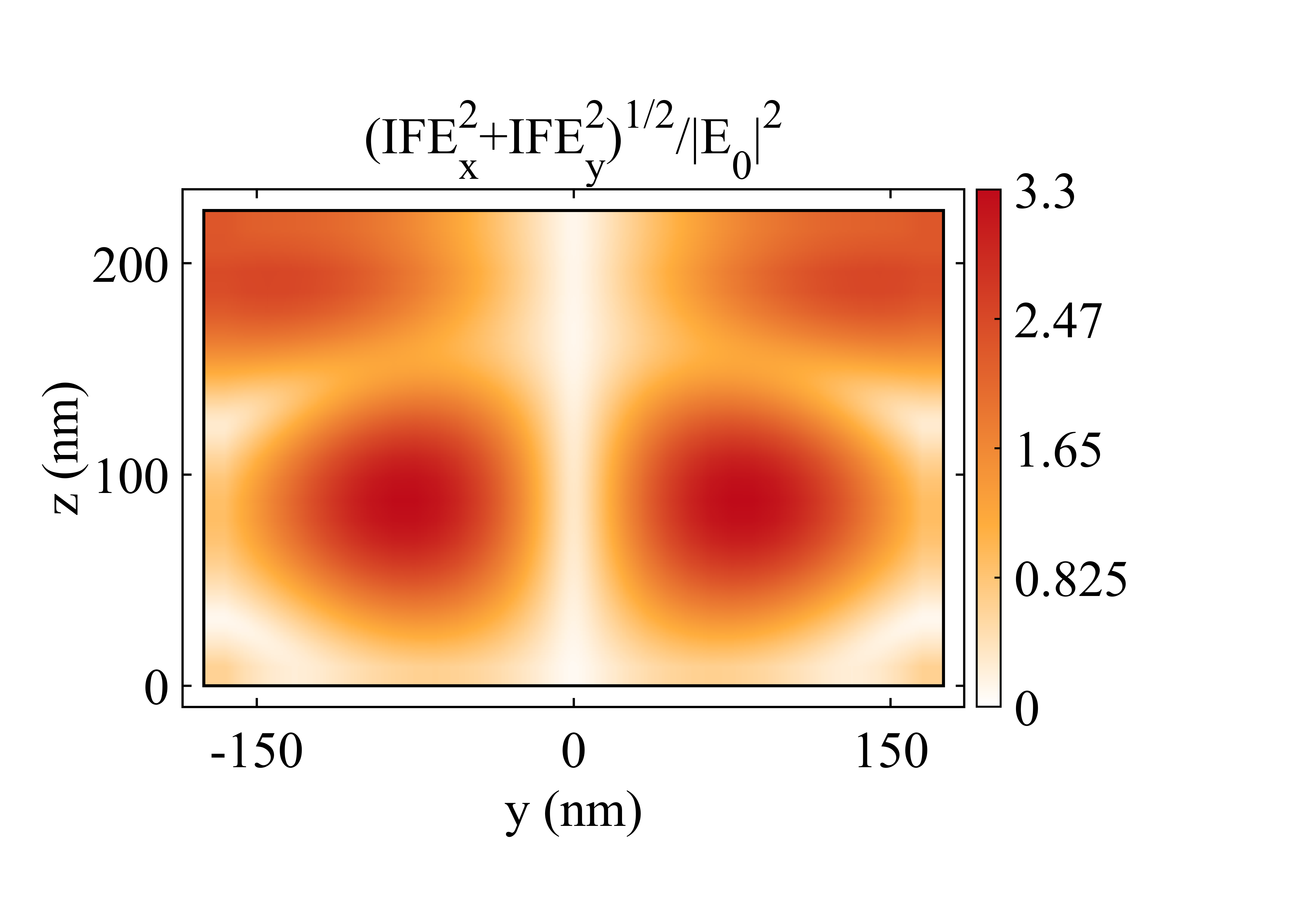}
\par\vspace{1.5mm}
(d)
\end{minipage}
\par\smallskip
\caption{Normalized in-plane IFE-related magnitude, $\sqrt{\mathrm{IFE}_x^2+\mathrm{IFE}_y^2}/|E_0|^2$, inside the BIG nanodisk on the SGG/BIG photonic-crystal substrate under circularly polarized excitation. Panels (a,c) correspond to $\lambda=837$~nm and panels (b,d) to $\lambda=849$~nm. Panels (a,b) show horizontal $xy$-cross-sections through the axial IFE hotspot ($z\simeq50$~nm in both panels), and panels (c,d) show vertical $yz$-cross-sections near the disk center. In the horizontal cross-sections, the white arrows show the direction of $(\mathrm{IFE}_x,\mathrm{IFE}_y)$; arrow lengths are scaled separately in each panel. The black contour marks the disk boundary. Note the different color scales.}
\label{fig:IFE_inplane_appendix}
\end{IFEplacedfigure}
\begin{multicols}{2}

\end{multicols}
\normalcolor
\begin{multicols}{2}

\end{multicols}

\begin{thebibliography}{42}
\providecommand{\natexlab}[1]{#1}
\providecommand{\url}[1]{\texttt{#1}}
\expandafter\ifx\csname urlstyle\endcsname\relax
  \providecommand{\doi}[1]{doi: #1}\else
  \providecommand{\doi}{doi: \begingroup \urlstyle{rm}\Url}\fi

\bibitem[\protect\citeauthoryear{Wang, Gao, Jiao, and Wang}{Wang
  et~al.}{2022}]{wang2022recent}
Kai Wang, Yong-Pan Gao, Rongzhen Jiao, and Chuan Wang.
\newblock ``Recent progress on optomagnetic coupling and optical manipulation
  based on cavity-optomagnonics.''
\newblock {\it Frontiers of Physics\/}~17, no. 4 (2022): 42201.

\bibitem[\protect\citeauthoryear{Zhang, Zhu, Zou, and Tang}{Zhang
  et~al.}{2016}]{zhang2016optomagnonic}
Xufeng Zhang, Na~Zhu, Chang-Ling Zou, and Hong~X Tang.
\newblock ``Optomagnonic whispering gallery microresonators.''
\newblock {\it Physical review letters\/}~117, no. 12 (2016): 123605.

\bibitem[\protect\citeauthoryear{Zhu, Zhang, Han, Zou, and Tang}{Zhu
  et~al.}{2022}]{zhu2022inverse}
Na~Zhu, Xufeng Zhang, Xu~Han, Chang-Ling Zou, and Hong~X Tang.
\newblock ``Inverse faraday effect in an optomagnonic waveguide.''
\newblock {\it Physical Review Applied\/}~18, no. 2 (2022): 024046.

\bibitem[\protect\citeauthoryear{Maccaferri, Zubritskaya, Razdolski, Chioar,
  Belotelov, Kapaklis, Oppeneer, and Dmitriev}{Maccaferri
  et~al.}{2020}]{maccaferri2020nanoscale}
Nicol{\`o} Maccaferri, Irina Zubritskaya, Ilya Razdolski, Ioan-Augustin Chioar,
  Vladimir Belotelov, Vassilios Kapaklis, Peter~M Oppeneer, and Alexandre
  Dmitriev.
\newblock ``Nanoscale magnetophotonics.''
\newblock {\it Journal of Applied Physics\/}~127, no. 8 (2020).

\bibitem[\protect\citeauthoryear{Voronov, Karki, Ignatyeva, Kozhaev, Levy, and
  Belotelov}{Voronov et~al.}{2020}]{voronov2020magneto}
Andrey~A Voronov, Dolendra Karki, Daria~O Ignatyeva, Mikhail~A Kozhaev, Miguel
  Levy, and Vladimir~I Belotelov.
\newblock ``Magneto-optics of subwavelength all-dielectric gratings.''
\newblock {\it Optics Express\/}~28, no. 12 (2020): 17988--17996.

\bibitem[\protect\citeauthoryear{Rameshti, Kusminskiy, Haigh, Usami,
  Lachance-Quirion, Nakamura, Hu, Tang, Bauer, and Blanter}{Rameshti
  et~al.}{2022}]{rameshti2022cavity}
Babak~Zare Rameshti, Silvia~Viola Kusminskiy, James~A Haigh, Koji Usami, Dany
  Lachance-Quirion, Yasunobu Nakamura, Can-Ming Hu, Hong~X Tang, Gerrit~EW
  Bauer, and Yaroslav~M Blanter.
\newblock ``Cavity magnonics.''
\newblock {\it Physics Reports\/}~979 (2022): 1--61.

\bibitem[\protect\citeauthoryear{Jesenska, Yoshida, Shinozaki, Ishibashi,
  Beran, Zahradnik, Antos, Ku{\v{c}}era, and Veis}{Jesenska
  et~al.}{2016}]{jesenska2016optical}
Eva Jesenska, Tomohiko Yoshida, Kenji Shinozaki, Takayuki Ishibashi, Lukas
  Beran, Martin Zahradnik, Roman Antos, Miroslav Ku{\v{c}}era, and Martin Veis.
\newblock ``Optical and magneto-optical properties of Bi substituted yttrium
  iron garnets prepared by metal organic decomposition.''
\newblock {\it Optical Materials Express\/}~6, no. 6 (2016): 1986--1997.

\bibitem[\protect\citeauthoryear{Sun, Du, Goto, Onbasli, Kim, Aimon, Hu, and
  Ross}{Sun et~al.}{2015}]{sun2015single}
Xue~Yin Sun, Qingyang Du, Taichi Goto, Mehmet~C Onbasli, Dong~Hun Kim,
  Nicolas~M Aimon, Juejun Hu, and Caroline~A Ross.
\newblock ``Single-step deposition of cerium-substituted yttrium iron garnet
  for monolithic on-chip optical isolation.''
\newblock {\it Acs Photonics\/}~2, no. 7 (2015): 856--863.

\bibitem[\protect\citeauthoryear{Kupchinskaya, Vetoshko, Kuzmichev, Polulyakh,
  Semuk, Pakhomov, Maksimov, Berzhansky, Belotelov, and Chernov}{Kupchinskaya
  et~al.}{2024}]{kupchinskaya2024magneto}
Nadezhda~E Kupchinskaya, Petr~M Vetoshko, Alexey~N Kuzmichev, Sergey~N
  Polulyakh, Eugeny~Y Semuk, Alexander~S Pakhomov, Fedor~M Maksimov, Vladimir~N
  Berzhansky, Vladimir~I Belotelov, and Alexander~I Chernov.
\newblock ``Magneto-optical epitaxial bismuth-substituted yttrium iron garnet
  thin films on a diamagnetic substrate for low temperature applications.''
\newblock {\it Journal of Magnetism and Magnetic Materials\/}~591 (2024):
  171623.

\bibitem[\protect\citeauthoryear{Legrand, Kemna, Sch{\"a}ren, Wang, Petrosyan,
  Holder, Schlitz, Aguirre, Lammel, and Gambardella}{Legrand
  et~al.}{2026}]{legrand2026lattice}
William Legrand, Yana Kemna, Stefan Sch{\"a}ren, Hanchen Wang, Davit Petrosyan,
  Luise Holder, Richard Schlitz, Myriam~H Aguirre, Michaela Lammel, and Pietro
  Gambardella.
\newblock ``Lattice-tunable substituted iron garnets for low-temperature
  magnonics.''
\newblock {\it Advanced Functional Materials\/}~36, no. 20 (2026): 2503644.

\bibitem[\protect\citeauthoryear{Flaj{\v{s}}man, Peeters, Kosunen, Yao,
  Lindfors-Vrejoiu, and Van~Dijken}{Flaj{\v{s}}man
  et~al.}{2026}]{flajvsman2026magnetic}
Luk{\'a}{\v{s}} Flaj{\v{s}}man, Lars Peeters, Armi Kosunen, Lide Yao, Ionela
  Lindfors-Vrejoiu, and Sebastiaan Van~Dijken.
\newblock ``Magnetic and structural properties of epitaxial Er-substituted
  yttrium iron garnet films grown by pulsed laser deposition.''
\newblock {\it Physical Review Materials\/}~10, no. 5 (2026): 054408.

\bibitem[\protect\citeauthoryear{Dubs, Surzhenko, Linke, Danilewsky,
  Br{\"u}ckner, and Dellith}{Dubs et~al.}{2017}]{dubs2017sub}
Carsten Dubs, Oleksii Surzhenko, Ralf Linke, Andreas Danilewsky, Uwe
  Br{\"u}ckner, and Jan Dellith.
\newblock ``Sub-micrometer yttrium iron garnet LPE films with low ferromagnetic
  resonance losses.''
\newblock {\it Journal of Physics D: Applied Physics\/}~50, no. 20 (2017):
  204005.

\bibitem[\protect\citeauthoryear{Hoefer, Kirby, and DeShazer}{Hoefer
  et~al.}{1988}]{hoefer1988thermo}
CS~Hoefer, KW~Kirby, and LG~DeShazer.
\newblock ``Thermo-optic properties of gadolinium garnet laser crystals.''
\newblock {\it Journal of the Optical Society of America B\/}~5, no. 11 (1988):
  2327--2332.

\bibitem[\protect\citeauthoryear{Kuzmichev, Vetoshko, Pavlyuk, Holin, Knyazev,
  Kaminskiy, Demirchyan, Tyumenev, Kalashnikov, Stolyarov, et~al.}{Kuzmichev
  et~al.}{2026}]{kuzmichev2026micrometer}
AN~Kuzmichev, PM~Vetoshko, EI~Pavlyuk, AA~Holin, GA~Knyazev, AS~Kaminskiy,
  SS~Demirchyan, R~Tyumenev, DS~Kalashnikov, VS~Stolyarov, et~al.
\newblock ``Micrometer thick single crystal iron-garnet films on a diamagnetic
  buffer layer for cryogenic applications.''
\newblock {\it Journal of Magnetism and Magnetic Materials\/} (2026): 174005.

\bibitem[\protect\citeauthoryear{Demirchyan, Krichevsky, and
  Belotelov}{Demirchyan et~al.}{2026}]{demirchyan2026magnon}
SS~Demirchyan, DM~Krichevsky, and VI~Belotelov.
\newblock ``Magnon--photon coupling in the YIG-based disk and ring
  microcavities.''
\newblock {\it Journal of Applied Physics\/}~139, no. 11 (2026).

\bibitem[\protect\citeauthoryear{Barsukova, Shorokhov, Musorin, Neshev,
  Kivshar, and Fedyanin}{Barsukova et~al.}{2017}]{barsukova2017magneto}
Maria~G Barsukova, Alexander~S Shorokhov, Alexander~I Musorin, Dragomir~N
  Neshev, Yuri~S Kivshar, and Andrey~A Fedyanin.
\newblock ``Magneto-optical response enhanced by Mie resonances in
  nanoantennas.''
\newblock {\it Acs Photonics\/}~4, no. 10 (2017): 2390--2395.

\bibitem[\protect\citeauthoryear{Marinchio, Carminati, Garcia-Martin, and
  Saenz}{Marinchio et~al.}{2014}]{marinchio2014magneto}
H.~Marinchio, R.~Carminati, A.~Garcia-Martin, and J.~J. Saenz.
\newblock ``Magneto-optical Kerr effect in resonant subwavelength nanowire
  gratings.''
\newblock {\it New Journal of Physics\/}~16, no. 1 (2014): 015007.

\bibitem[\protect\citeauthoryear{Gamet, Varghese, Verrier, and Royer}{Gamet
  et~al.}{2017}]{gamet2017enhancement}
E.~Gamet, B.~Varghese, I.~Verrier, and F.~Royer.
\newblock ``Enhancement of magneto-optical effects by a single 1D all
  dielectric resonant grating.''
\newblock {\it Journal of Physics D: Applied Physics\/}~50, no. 49 (2017):
  495105.

\bibitem[\protect\citeauthoryear{Christofi, Kawaguchi, Al{\`u}, and
  Khanikaev}{Christofi et~al.}{2018}]{christofi2018giant}
Aristi Christofi, Yuma Kawaguchi, Andrea Al{\`u}, and Alexander~B. Khanikaev.
\newblock ``Giant enhancement of Faraday rotation due to electromagnetically
  induced transparency in all-dielectric magneto-optical metasurfaces.''
\newblock {\it Optics Letters\/}~43, no. 8 (2018): 1838--1841.

\bibitem[\protect\citeauthoryear{Zorina, Ignatyeva, Bezmenova, Kalish, Xia, Bi,
  and Belotelov}{Zorina et~al.}{2026}]{zorina2026enhanced}
P.~V. Zorina, D.~O. Ignatyeva, A.~E. Bezmenova, A.~N. Kalish, S.~Xia, L.~Bi,
  and V.~I. Belotelov. 2026.
\newblock ``Enhanced magneto-optical intensity effect in a helicity-preserving
  all-dielectric metasurface at Mie resonances and the anapole state.'' , .

\bibitem[\protect\citeauthoryear{Chernyak, Barsukova, Shorokhov, Musorin, and
  Fedyanin}{Chernyak et~al.}{2020}]{chernyak2020bound}
A.~M. Chernyak, M.~G. Barsukova, A.~S. Shorokhov, A.~I. Musorin, and A.~A.
  Fedyanin.
\newblock ``Bound States in the Continuum in Magnetophotonic Metasurfaces.''
\newblock {\it JETP Letters\/}~111, no. 1 (2020): 46--49.

\bibitem[\protect\citeauthoryear{Zakharov and Poddubny}{Zakharov and
  Poddubny}{2020}]{zakharov2020transverse}
V.~A. Zakharov and A.~N. Poddubny.
\newblock ``Transverse magneto-optical Kerr effect enhanced at the bound states
  in the continuum.''
\newblock {\it Physical Review A\/}~101, no. 4 (2020): 043848.

\bibitem[\protect\citeauthoryear{Mamian, Popov, Frolov, and Fedyanin}{Mamian
  et~al.}{2025}]{mamian2025tailoring}
Karen~A Mamian, Vladimir~V Popov, A~Yu Frolov, and Andrey~A Fedyanin.
\newblock ``Tailoring transverse magneto-optical Kerr effect enhancement in
  Mie-resonant nanowire-based metasurfaces.''
\newblock {\it Applied Physics Letters\/}~126, no. 8 (2025).

\bibitem[\protect\citeauthoryear{Krichevsky, Ignatyeva, and
  Belotelov}{Krichevsky et~al.}{2024}]{krichevsky2024inverse}
Denis~M Krichevsky, Daria~O Ignatyeva, and Vladimir~I Belotelov.
\newblock ``Inverse Faraday effect at Mie resonances.''
\newblock {\it Physical Review Applied\/}~22, no. 6 (2024): 064087.

\bibitem[\protect\citeauthoryear{Krichevsky, Ozerov, Bel’kova, Sylgacheva,
  Kalish, Evstigneeva, Pakhomov, Mikhailova, Lyashko, Kudryashov,
  et~al.}{Krichevsky et~al.}{2024}]{krichevsky2024spatially}
Denis~M Krichevsky, Vladislav~A Ozerov, Alexandra~V Bel’kova, Daria~A
  Sylgacheva, Andrey~N Kalish, Svetlana~A Evstigneeva, Alexander~S Pakhomov,
  Tatiana~V Mikhailova, Sergey~D Lyashko, Alexander~L Kudryashov, et~al.
\newblock ``Spatially inhomogeneous inverse Faraday effect provides tunable
  nonthermal excitation of exchange dominated spin waves.''
\newblock {\it Nanophotonics\/}~13, no. 3 (2024): 299--306.

\bibitem[\protect\citeauthoryear{Hayashi, Dao, Gross, Ranno, Sia, Fakhrul, Du,
  Chatterjee, Hu, and Ross}{Hayashi et~al.}{2024}]{hayashi2024magneto}
Kensuke Hayashi, Khoi~Phuong Dao, Miela~J Gross, Luigi Ranno, Jia Xu~Brian Sia,
  Takian Fakhrul, Qingyang Du, Nilanjan Chatterjee, Juejun Hu, and Caroline~A
  Ross.
\newblock ``Magneto-Optical Bi-Substituted Yttrium and Terbium Iron Garnets for
  On-Chip Crystallization via Microheaters.''
\newblock {\it Advanced Optical Materials\/}~12, no. 25 (2024): 2400708.

\bibitem[\protect\citeauthoryear{Kolodny and Iorsh}{Kolodny and
  Iorsh}{2019}]{kolodny2019q}
Stanislav Kolodny and Ivan Iorsh.
\newblock ``Q/V enhancement of micropillar resonator in bound states in the
  continuum regime.''
\newblock {\it Optics Letters\/}~45, no. 1 (2019): 181--183.

\bibitem[\protect\citeauthoryear{Yang, Xia, Zhang, Wang, Wei, Wang, Chen, Zhou,
  Qin, Hu, et~al.}{Yang et~al.}{2025}]{yang2025giant}
Weihao Yang, Shuang Xia, Tianchi Zhang, Jialong Wang, Zixuan Wei, Hanbin Wang,
  Yiqing Chen, Yuting Zhou, Jun Qin, Aimin Hu, et~al.
\newblock ``Giant Faraday rotation in high-Q all-dielectric magneto-optical
  metasurfaces embedded in photonic crystal cavities.''
\newblock {\it Photonics Research\/}~13, no. 12 (2025): 3499--3510.

\bibitem[\protect\citeauthoryear{Kahl and Grishin}{Kahl and
  Grishin}{2004}]{kahl2004enhanced}
Soren Kahl and Alexander~M Grishin.
\newblock ``Enhanced Faraday rotation in all-garnet magneto-optical photonic
  crystal.''
\newblock {\it Applied Physics Letters\/}~84, no. 9 (2004): 1438--1440.

\bibitem[\protect\citeauthoryear{Khartsev and Grishin}{Khartsev and
  Grishin}{2007}]{khartsev2007high}
S~Khartsev and A~Grishin.
\newblock ``High performance magneto-optical photonic crystals.''
\newblock {\it J. Appl. Phys\/}~101 (2007): 053--906.

\bibitem[\protect\citeauthoryear{Berzhansky, Shaposhnikov, Prokopov,
  Karavainikov, Mikhailova, Semuk, Sharipova, Dolgova, Fedyanin, Kotov,
  et~al.}{Berzhansky et~al.}{2011}]{berzhansky2011one}
VN~Berzhansky, AN~Shaposhnikov, AR~Prokopov, AV~Karavainikov, TV~Mikhailova,
  E~Yu Semuk, MI~Sharipova, TV~Dolgova, AA~Fedyanin, VA~Kotov, et~al.
\newblock ``One-dimensional magnetophotonic crystals based on double-layer
  Bi-substituted iron garnet films.''
\newblock {\it Materialwissenschaft und Werkstofftechnik\/}~42, no. 1 (2011):
  19--23.

\bibitem[\protect\citeauthoryear{Ignatyeva, Mikhailova, Kapralov, Lyashko,
  Berzhansky, and Belotelov}{Ignatyeva et~al.}{2024}]{ignatyeva2024asymmetric}
DO~Ignatyeva, TV~Mikhailova, PO~Kapralov, SD~Lyashko, VN~Berzhansky, and
  VI~Belotelov.
\newblock ``Asymmetric Faraday effect caused by a break of spatial symmetry.''
\newblock {\it Physical Review Applied\/}~22, no. 4 (2024): 044064.

\bibitem[\protect\citeauthoryear{Dzibrou and Grishin}{Dzibrou and
  Grishin}{2009}]{dzibrou2009fitting}
DO~Dzibrou and AM~Grishin.
\newblock ``Fitting transmission and Faraday rotation spectra of
  [Bi3Fe5O12/Sm3Ga5O12] m magneto-optical photonic crystals.''
\newblock {\it Journal of Applied Physics\/}~106, no. 4 (2009).

\bibitem[\protect\citeauthoryear{Khartsev and Grishin}{Khartsev and
  Grishin}{2011}]{khartsev2011high}
Sergey~I Khartsev and Alexander~M Grishin.
\newblock ``High performance latching-type luminescent magneto-optical photonic
  crystals.''
\newblock {\it Optics letters\/}~36, no. 15 (2011): 2806--2808.

\bibitem[\protect\citeauthoryear{Alaee, Rockstuhl, and
  Fernandez-Corbaton}{Alaee et~al.}{2018}]{alaee2018electromagnetic}
Rasoul Alaee, Carsten Rockstuhl, and Ivan Fernandez-Corbaton.
\newblock ``An electromagnetic multipole expansion beyond the long-wavelength
  approximation.''
\newblock {\it Optics Communications\/}~407 (2018): 17--21.

\bibitem[\protect\citeauthoryear{Pitaevskii}{Pitaevskii}{1961}]{pitaevskii1961electric}
LP~Pitaevskii.
\newblock ``Electric forces in a transparent dispersive medium.''
\newblock {\it Sov. Phys. JETP\/}~12, no. 5 (1961): 1008--1013.

\bibitem[\protect\citeauthoryear{Pershan, Van~der Ziel, and Malmstrom}{Pershan
  et~al.}{1966}]{pershan1966theoretical}
PS~Pershan, JP~Van~der Ziel, and LD~Malmstrom.
\newblock ``Theoretical discussion of the inverse Faraday effect, Raman
  scattering, and related phenomena.''
\newblock {\it Physical review\/}~143, no. 2 (1966): 574.

\bibitem[\protect\citeauthoryear{Kimel, Kirilyuk, Usachev, Pisarev, Balbashov,
  and Rasing}{Kimel et~al.}{2005}]{kimel2005ultrafast}
AV~Kimel, A~Kirilyuk, PA~Usachev, RV~Pisarev, AM~Balbashov, and Th~Rasing.
\newblock ``Ultrafast non-thermal control of magnetization by instantaneous
  photomagnetic pulses.''
\newblock {\it Nature\/}~435, no. 7042 (2005): 655--657.

\bibitem[\protect\citeauthoryear{Hansteen, Kimel, Kirilyuk, and
  Rasing}{Hansteen et~al.}{2006}]{hansteen2006nonthermal}
Fredrik Hansteen, Alexey Kimel, Andrei Kirilyuk, and Theo Rasing.
\newblock ``Nonthermal ultrafast optical control of the magnetization in garnet
  films.''
\newblock {\it Physical Review B—Condensed Matter and Materials
  Physics\/}~73, no. 1 (2006): 014421.

\bibitem[\protect\citeauthoryear{Stanciu, Hansteen, Kimel, Kirilyuk, Tsukamoto,
  Itoh, and Rasing}{Stanciu et~al.}{2007}]{stanciu2007all}
Claudiu~D Stanciu, Fredrik Hansteen, Alexey~V Kimel, Andrei Kirilyuk, Arata
  Tsukamoto, Achioshi Itoh, and Th~Rasing.
\newblock ``All-optical magnetic recording with circularly polarized light.''
\newblock {\it Physical review letters\/}~99, no. 4 (2007): 047601.

\bibitem[\protect\citeauthoryear{Stupakiewicz, Szerenos, Davydova, Zvezdin,
  Zvezdin, Kirilyuk, and Kimel}{Stupakiewicz
  et~al.}{2019}]{stupakiewicz2019selection}
A~Stupakiewicz, Krzysztof Szerenos, MD~Davydova, KA~Zvezdin, AK~Zvezdin, Andrei
  Kirilyuk, and AV~Kimel.
\newblock ``Selection rules for all-optical magnetic recording in iron
  garnet.''
\newblock {\it Nature communications\/}~10, no. 1 (2019): 612.

\bibitem[\protect\citeauthoryear{Kozhaev, Chernov, Sylgacheva, Shaposhnikov,
  Prokopov, Berzhansky, Zvezdin, and Belotelov}{Kozhaev
  et~al.}{2018}]{kozhaev2018giant}
Mikhail~A Kozhaev, Alexander~I Chernov, Daria~A Sylgacheva, Alexander~N
  Shaposhnikov, Anatoly~R Prokopov, Vladimir~N Berzhansky, Anatoly~K Zvezdin,
  and Vladimir~I Belotelov.
\newblock ``Giant peak of the Inverse Faraday effect in the band gap of
  magnetophotonic microcavity.''
\newblock {\it Scientific reports\/}~8, no. 1 (2018): 11435.

\end{thebibliography}
\end{document}